\documentclass[11pt,a4paper]{article}

\usepackage[utf8]{inputenc}
\usepackage[english]{babel}

\usepackage{geometry}
\usepackage{graphicx}
\usepackage{amssymb,amsmath,amsthm}
\usepackage{stmaryrd}
\usepackage{delarray}
\usepackage{enumitem}
\usepackage{tikz}
\usepackage{hyperref}
\graphicspath{{figures/}}
\usepackage{authblk}
\usepackage{wrapfig}
\usepackage{multicol}
\usepackage{float}
\usepackage{appendix}
\usepackage{pdfpages}
\usepackage{tikz}

\usetikzlibrary{calc,decorations.pathreplacing,patterns}

\newtheorem{theorem}{\textbf{Theorem}}
\newtheorem*{theorem*}{\textbf{Theorem}}
\newtheorem{lemma}[theorem]{\textbf{Lemma}}
\newtheorem{proposition}[theorem]{\textbf{Proposition}}

\newtheorem{corollary}[theorem]{\textbf{Corollary}}
\newtheorem{remark}{\textbf{Remark}}

\newtheorem{definition}{\textbf{Definition}}

\newtheorem{remark*}{Remark}

\newtheorem{innercustomtheorem}{Theorem}
\newenvironment{customtheorem}[1]
  {\renewcommand\theinnercustomtheorem{#1}\innercustomtheorem}
  {\endinnercustomtheorem}

\renewcommand{\int}[1]{[#1]}
\newcommand{\intz}[1]{\llbracket#1\rrbracket}
\newcommand{\N}{\mathbb{N}}

\renewcommand{\O}{\mathcal{O}}

\newcommand{\Poly}{{\mathsf{P}}}
\newcommand{\NP}{{\mathsf{NP}}}
\newcommand{\coNP}{{\mathsf{coNP}}}

\newcommand{\soo}[1]{\smash{\overline{\overline{#1}}}}

\newcommand{\akb}[2]{{\mathcal{P}(#1,#2)}}
\newcommand{\akbq}[3]{{\mathcal{P}_{#3}(#1,#2)}}
\newcommand{\bool}{\{0,1\}}

\newcommand{\dyna}[1]{\mathcal{G}_{#1}}

\newcommand*{\constant}[1]{\mathrm{constant}_{#1}}

\newcommand*{\recolor}[1]{\mathrm{recolor}_{#1}}
\newcommand*{\join}[1]{\mathrm{join}_{#1}}

\newcommand{\decisionpbw}[4]{
  \fbox{\parbox{{#4}\textwidth}{{\bf {#1}}\\{\it Input:} {#2}\\{\it Question:} {#3}}}
}

\usepackage[textsize=small]{todonotes}

\title{Rice-like complexity lower bounds for\\Boolean and uniform automata networks}
\author[a]{Ali\'enor Goubault--Larrecq}
\author[a,b]{K\'evin Perrot}
\affil[a]{Aix-Marseille Univ., Univ. de Toulon, CNRS, LIS, Marseille, France}
\affil[b]{Universit\'e publique, France}
\date{}

\begin{document} 
	%%%%%%%%%%%%%%%%%%%%%%%%%%%%%%%%
	\maketitle
	
	\begin{abstract}
          Automata networks are a versatile model of finite discrete dynamical systems
          composed of interacting entities (the automata),
          able to embed any directed graph as a dynamics on its space of configurations (the set of vertices,
          representing all the assignments of a state to each entity).
          In this world, virtually any question is decidable by a simple exhaustive search.
          We lever the Rice-like complexity lower bound, stating that any non-trivial monadic second order logic
          question on the graph of its dynamics
          is $\NP$-hard or $\coNP$-hard (given the automata network description),
          to bounded alphabets (including the Boolean case).
          This restriction is particularly meaningful for applications to ``complex systems'',
          where each entity has a restricted set of possible states (its alphabet).
          For the deterministic case, trivial questions are solvable in constant time, hence there is a sharp gap
          in complexity for the algorithmic solving of concrete problems on them.
          For the non-deterministic case, non-triviality is defined at bounded cliquewidth,
          which offers a structure to establish metatheorems of complexity lower bounds.
	\end{abstract}

	%%%%%%%%%%%%%%%%%%%%%%%%%%%%%%%%
        \pagebreak
	\tableofcontents
        \pagebreak

	%%%%%%%%%%%%%%%%%%%%%%%%%%%%%%%%
	\section{Introduction}

        A natural way to formalize the intuitive notion of ``complexity'' arising in the dynamics of
        discrete dynamical systems employs the well established theory of algorithmic complexity.
        This approach has notably been carried on cellular automata
        \cite{s89,Griffeath1996LifeWD,Moore1997,woods110,formenti2017},
        lattice gas
        \cite{moore1997predicting,Machta1997}
        and sandpile
        \cite{1999-Moore-complexitySandpiles,fp19},
        regarding the prediction of their dynamics.
        It has also been introduced on automata networks,
        a bioinspired model of computation on which the present work is grounded,
        with a particular focus on fixed points
        \cite{a85,fo89,o92,k08}.
        %Introduction: algorithmic complexity of ``complex systems''
        %Examples on different models:
        %- Sutner finite CA~\cite{s89}
        %- Life Without Death is P-complete \cite{Griffeath1996LifeWD},
        %- Majority-Vote Cellular Automata, Ising Dynamics, and P-Completeness \cite{Moore1997},
        %- P-completeness of Cellular Automaton Rule 110 \cite{woods110}
        %- Computational complexity of finite asynchronous cellular automata \cite{formenti2017}
        %- Predicting lattice gases is P-complete \cite{moore1997predicting},
        %- The computational complexity of the Lorentz lattice gas \cite{Machta1997},
        %- Sandpiles (Moore Nilsen) \cite{1999-Moore-complexitySandpiles},
        %- Sandpiles survey \cite{fp19}
        %Finish with ANs:
        %- Alon, Floréen, Orponen, Kosub on fixed points \cite{a85,fo89,o92,k08},

        Automata networks are a general model of interacting entities,
        where each \emph{automaton} holds a state (taken among a finite alphabet),
        and updates it according to each other's current state.
        In the deterministic setting,
        the behavior of each automaton is described by a local update function,
        which are all applied synchronously in discrete steps.
        In the non-deterministic setting,
        multiple concurrent behaviors are possible.
        The set of automata and states are finite, so is the configuration space
        (assigning a state to each automaton).
        The \emph{dynamics} of an automata network is the graph of the transition function or relation
        on its configuration space.
        This model is very general, in the sense that any directed graph
        is the dynamics of some automata network (the graph has out-degree $1$ in the deterministic setting).
        Automata networks have applications in many fields,
        most notably for the modeling of gene regulation mechanisms and biological systems in general
        \cite{k69,t73,lr23}.
        In this context, restricting the set of possible states of each automaton
        is particularly meaningful
        \cite{e04,ks08,m98,t95,lr23}.
        A prototypical example is the Boolean case,
        where each automaton holds a state among $\{0,1\}$.
        This is the constraint targeted in the present article.
        The consideration of alternative update modes is another central concern in the community
        (see~\cite{ps22} for a survey),
        here we stick to the synchronous case, also called the parallel update mode.
        %Automate networks are a general model of interacting entities
        %(any directed graph as dynamics, of out-degree $1$ in the deterministic setting).
        %Gene regulatory networks (Kauffman, Thomas), update modes, etc...
        %importance of restricting the alphabet size (arabidopsis Boolean).

        A pillar of computer science is Rice theorem~\cite{r53}:
        any non-trivial semantic property of programs is undecidable.
        It is striking for its generality and the sharp dichotomy of difficulty
        between trivial and non-trivial problems.
        We look for analogs in this spirit.
        A finite discrete dynamical system can only raise decidable questions (simply because it is finite),
        hence we shift the perspective from computability to complexity theory.
        We lever the Rice-like complexity lower bounds of~\cite{ggpt21, gglgopt25},
        from succinct graph representations (where it contrasts Courcelle theorem)
        to natural models of interacting entities (automata networks on bounded alphabets).
        Such metatheorems are obtained using the expressiveness of graph logics,
        all at monadic second order (MSO) in the present work.
        A key consists in defining a notion of non-triviality as general as possible.
        It turns out to be a sharp and deep dichotomy in the deterministic case (AN) because trivial
        questions are answered in constant time.
        In the non-deterministic case (NAN), we employ a notion of non-triviality relative to
        dynamics of bounded cliquewidth, which we call cw-non-trivial
        (its necessity is addressed in~\cite{gglgopt25} and discussed in the perspectives).
        Our results apply to uniform automata networks where
        all automata are constrained to have the same alphabet size (denoted $q$).
        All the necessary concepts will be defined in the preliminary section.
        We gather our main results in one statement, where in bold is the problem
        consisting in deciding whether the dynamics of an automata network
        (given a description of the behavior of its entities)
        verifies some property $\psi$ expressed in graph logics.

        %We love Rice :
        %Any non-trivial semantic property of programs is undecidable.
        %Here everything is finite therefore decidable :
        %shift from computability to complexity theory
        %We lever the Rice-like complexity lower bounds of~\cite{ggpt21,gamard2023hardness}
        %from succinct graph representations (counterpoint to Courcelle's theorem)
        %to natural models of interacting entities (uniform alphabets

        %Statement of our main result (MSO for uniform, both det and non-det):
        %need definitions (give them in English)

	\begin{theorem}
		\label{theorem:hard}
                Let $q\geq 2$ be any alphabet size for \underline{\smash{uniform}} automata networks.\\
                \underline{Deterministic:}\\
                %\begin{itemize}[nosep]
                  %\item
                    \textnormal{\textbf{a.}}
                    For any non-trivial MSO formula,
                    {\bf $\psi$-AN-dynamics} is $\NP$- or $\coNP$-hard.\\
                  %\item 
                    \textnormal{\textbf{b.}}
                    For any $q$-non-trivial MSO formula,
                    {\bf $\psi$-$q$-AN-dynamics} is $\NP$- or $\coNP$-hard.\\
                %\end{itemize}
                \underline{Non-deterministic:}\\
                %\begin{itemize}[nosep]
                  %\item
                    \textnormal{\textbf{c.}}
                    For any cw-non-trivial MSO formula,
                    {\bf $\psi$-NAN-dynamics} is $\NP$- or $\coNP$-hard.\\
                  %\item
                    \textnormal{\textbf{d.}}
                    For any $q$-cw-non-trivial MSO formula,
                    {\bf $\psi$-$q$-NAN-dynamics} is $\NP$- or $\coNP$-hard.
                %\end{itemize}
	\end{theorem}
 
        %Argue that $q$-non-trivial and $q$-cw-non-trivial are necessary.
        %Deterministic gives a sharp complexity dichotomy with $\O(1)$,
        %but we do not have this for non-det (other parameters are investigated, to develop in perspectives).

        Part~c of Theorem~\ref{theorem:hard} is proven in~\cite{gglgopt25}.
        The $\NP$ and  $\coNP$ symmetry is necessary in such a general statement,
        because some problems expressible at first order (such as the existence of a fixed point in the dynamics)
        are known to be $\NP$-complete~\cite{a85},
        and the graph logics we consider are closed by complementation (simply by adding a negation on top of the formula).
        The proof technique nevertheless gives a clear characterization of which of these
        two lower bounds is proven for each formula $\psi$,
        through the concept of saturating graph (denoted $\Omega_m$ because it only depends on the quantifier
        rank $m$ of $\psi$) introduced in~\cite{gglgopt25}.
        If $\Omega_m$ satisfies an MSO formula $\psi$, then the corresponding \textbf{$\psi$-dynamics} problem
        is proven to be $\NP$-hard, otherwise it is proven to be $\coNP$-hard.
        %Argue that the $\NP$/$\coNP$ symmetry is necessary.
        %A way to know which way: when the saturating graph is a model (resp. a counter-model)
        %then \textbf{$\psi$-dynamics} is proven to be $\NP$-hard (resp. $\coNP$-hard).
	
        In Section~\ref{section:prelim} we introduce all the necessary notions involved in the statement
        and in the proof of Theorem~\ref{theorem:hard}, including graph logics and clique-decompositions.
        Theorem~\ref{theorem:hard} extends known results in two directions:
        \begin{itemize}[nosep]
          \item generalizing to MSO the result of~\cite{ggpt21} in the deterministic case,
            by employing the abstract pumping technique based on finite model theory from~\cite{gglgopt25},
          \item proving that the Rice-like complexity lower bounds also hold on $q$-uniform networks,
            where automata states are taken among a common alphabet of size $q\geq 2$.
        \end{itemize}
        The proof technique developed in~\cite{gglgopt25},
        on which our extensions are grounded, is detailed in Section~\ref{section:previous}
        where all the ingredients to design a polytime reduction from \textbf{SAT} are exposed.
        In Section~\ref{section:det} we apply the technique to deterministic dynamics,
        \emph{i.e.} to graphs of out-degree $1$, and prove Part~a of Theorem~\ref{theorem:hard}.
        Section~\ref{section:lowerbounds} is dedicated to the study of arithmetical considerations
        in order to obtain $q$-uniform dynamics (on a configuration space of size $q^n$
        for some number $n$ of automata) when pumping a part of size $a$ from a graph of size $b$
        (that is, obtaining potential dynamics of size $ak+b$ where $k$ is the number of copies
        of size $a$ which have been added to the initial graph of size $b$).
        We give a separate consideration to the Boolean case $q=2$,
        which is simpler and gives the intuitions for the last ingredients of our main results.
        In Section~\ref{section:qunif} we apply the arithmetics of the previous section
        to derive Parts~b and~d of Theorem~\ref{theorem:hard}
        which, to our point of view, are the most impacting new results presented in this paper,
        already asked for at the end of~\cite{ggpt21}.
        We conclude and present perspectives on how to pursue the quest for metatheorems
        on the complexity of discrete dynamical systems in Section~\ref{section:perspectives}.%
        \footnote{%
          This work is an update of~\cite{glp25a} (from v1 to v2)
          in order to bring the results from dynamics of bounded treewidth to dynamics of bounded cliquewidth,
          according to a recent update of \cite{gglgopt25} performing this extension of the results (from v2 to v3).%
        }

	%%%%%%%%%%%%%%%%%%%%%%%%%%%%%%%%
	\section{Preliminaries}
        \label{section:prelim}

	For $n,q \in \N_+$, we denote 
	$\int{n} = \{1,\dots,n\}$ and $\intz{q}=\{0,\dots,q-1\}$.
        Given a directed graph (digraph) $G = (V(G), E(G))$, we consider its size as  
        $|G| = |V(G)|$.
	Given a vertex $v \in V(G)$, the out-neighbors of $v$ are denoted 
        $G(v)=\{u \mid (v,u) \in E(G)\}$,
        and $|G(v)|$ is the out-degree of $v$.
	We say that a graph $G$ is of out-degree (exactly) $d \in \N$ when all its vertices have
	out-degree $d$.
	%$G$  is said \emph{connected} if for every pair of vertices $u, v \in V(G)$ there 
	%exists a path between $u$ and $v$ in the underlying undirected graph of $G$. A 
	%\emph{connected component} is a maximal connected subgraph and every graph 
	%can be partitioned into its set of connected components.  
	
	%%%%%%%%%%%%%%%%
	\subsection{Automata networks}

        A \emph{deterministic automata network} (AN) of size $n$ is a 
	function $f : X \to X$ where $X= \prod_{i \in \int{n}} A_i$ is the set of \emph{configurations} 
        of the system, and $A_i=\intz{q_i}$ is the set of \emph{states} of the $i^{\textnormal{th}}$ \emph{automaton} 
	of the network.
	The function $f$ can be split into \emph{local functions} 
	$\{f_i\}_{i \in \int{n}}$, where $\forall i \in [n], f_i: X \to A_i$. 
        Hence $f_i$ returns the state of the $i^{\textnormal{th}}$ automaton at the next step. 
	We can also retrieve $f$ from all the local functions since $\forall x \in X,
	f(x) = (f_1(x), f_2(x), ..., f_n(x))$. 
	An AN $f$ can be represented by its \emph{dynamics} or \emph{transition digraph} $\dyna{f}$,
	such that $V(\dyna{f}) = X$ and $E(\dyna{f}) = \{(x, f(x)) \mid \forall x \in V(\dyna{f})\}$. 
	This dynamics is deterministic, hence $\dyna{f}$ is a digraph
        having out-degree $1$ on each vertex, also called a \emph{functional digraph}
        (namely, the graph of the function $f$).

        A \emph{non-deterministic automata network} (NAN) of size $n$
        is a relation $f\subseteq X\times X$ where $f(x)$
        is the set of possible transitions from configuration $x\in X$
        (again with $X= \prod_{i \in \int{n}} A_i$).
        Remark that defining local relations $(f_i\subseteq X\times A_i)_{i\in\int{n}}$
        is not always possible, because it would suggest that
        $f(x)=\{y \mid \forall i\in\int{n},y_i\in f_i(x)\}$
        but this is not possible for all relations $f$ when $n>1$
        (and the number of automata is at the heart of our considerations,
        in particular regarding $q$-uniformity).
        The non-determinism we consider in this paper is global, by contrast with local non-determinism
        defined through local relations, where the image of a configuration $x$ is
        taken among the Cartesian product of the possibilities for each automaton
        ($f(x)\in f_1(x)\times f_2(x)\times\dots\times f_n(x)$).
        %For example, for $n=2$ with $A_1=A_2=\{0,1\}$
        %the NAN $f=\{(00,00),(00,11)\}$ cannot be decomposed.

        When all the state sets $A_i$ have the same size $q$, we say that $f$ is a
        \emph{$q$-uniform (non-deterministic) automata network} ($q$-AN or $q$-NAN).
        For $q=2$ we have $X = \{0, 1\}^n$ and we call $f$ a
        \emph{Boolean (non-deterministic) automata network} (BAN or BNAN).

        \begin{figure}
          \centering
          \includegraphics{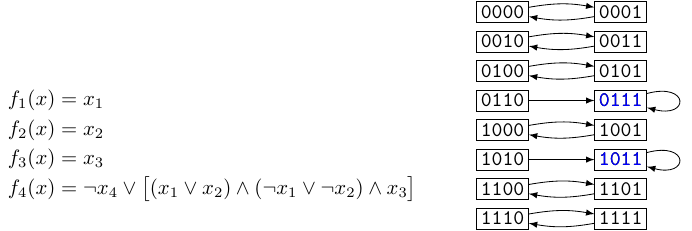}
          \caption{
            Example deterministic Boolean automata network of size $n=4$.
            Local functions $f_i:\{0,1\}^4\to\{0,1\}$ for $i\in\int{n}$ (left)
            and transition digraph dynamics $\dyna{f}$ on configuration space $\{0,1\}^4$ (right).
          }
          \label{fig:AN}
        \end{figure}

        \begin{figure}
          \centering
          \includegraphics{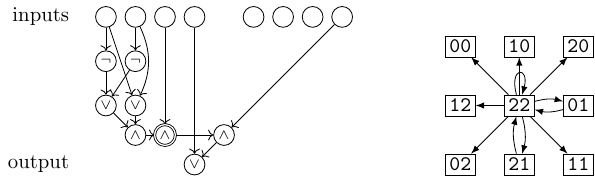}
          \caption{
            Example non-deterministic $q$-uniform automata network of size $n=2$
            on alphabet $\intz{q}=\{0,1,2\}$ with $q=3$.
            Circuit of the relation $f\subseteq\intz{q}^n\times\intz{q}^n$
            (left)
            and transition digraph dynamics $\dyna{f}$ on configuration space $\intz{q}^n$ (right).
            A configuration is encoded on four bits, where inputs $0000$ to $0001$ correspond
            respectively to configurations $00$ to $22$.
            The doubly circled node represents the
            evaluation of propositional formula $S=(x_1\vee x_2)\wedge(\neg x_1\vee\neg x_2)\wedge x_3$
            on the three first bits of the circuit's input.
            % strongly connected iff S is a tautology
          }
          \label{fig:NAN}
        \end{figure}

        \begin{remark}\label{rem:allG}
          For all graph $G$, there exists a NAN $f$
          such that $\dyna{f}=G$ (up to a renaming of the vertices).
          If $G$ has out-degree $1$, then there is also such an AN $f$.
          In particular, we can set $n=1$ automaton with state set $A_1=V(G)$.
          It also holds for $q$-uniform networks, provided the graph $G$ has size $q^n$ for some integer $n$.
        \end{remark}

        Given our focus on computational complexity, it is important to make explicit how ANs and NANs are encoded
        when given as inputs to problems.
        An AN $f$ is encoded as one Boolean circuit with
        $\lceil\log_2 |X|\rceil$ input bits
        and $\lceil\log_2 |X|\rceil$ output bits.
        The state set sizes $(|A_i|)_{i\in\int{n}}$ are also part of the input
        in order to encode/decode configurations as bitstrings,
        but formally we only require to provide the global size $N=|X|$.
        Given a configuration $x\in X$ as input to the circuit,
        its image $f(x)$ is read as the output taken modulo $N$
        (to avoid the $\coNP$-complete problem
        of checking that the circuit always outputs a bitstring from $0$ to $N-1$).
        In the Boolean case, each input/output bit of the circuit is the state of an automaton
        (circuits generalize propositional formulas,
        but all our results still hold for an encoding through formulas).
        A NAN $f$ is encoded as one Boolean circuit with
        $2\cdot\lceil\log_2 |X|\rceil$ input bits
        and $1$ output bit, indicating whether the two input configurations $x$ and $y$ verify $y\in f(x)$.
        The circuit size of an AN is restricted to some $\O(|X|\log_2|X|)$
        (the size of a lookup table for $\dyna{f}$ of out-degree $1$),
        and that of a NAN to some $\O(|X|^2)$
        (the size of an adjacency matrix for $\dyna{f}$).
        These encodings are also called \emph{succinct graph representations} of $\dyna{f}$~\cite{gw83}.
        It is important to precise the size $N=|X|$ of the configuration space when there is not restriction
        on the alphabets, but for $q$-uniform networks it is enough to give the number $n$ of automata
        (then we deduce that there are $q^n$ configurations).
        When constructing ANs or NANs in our proofs,
        it will always be straightforward to convert (in polynomial time)
        our descriptions of dynamics into local functions and their encodings as circuits.
        Examples of AN and NAN are given on Figures~\ref{fig:AN} and~\ref{fig:NAN}, respectively.
	
	%%%%%%%%%%%%%%%%
        \subsection{Graph logics}
	
	If $P$ is a property that automata networks may or may not satisfy, and $f$ is
	an AN or a NAN, then we write $f \vDash P$ if $f$ satisfies $P$, and $f
	\nvDash P$ otherwise. We say that $f$ is a \emph{model} of $P$ in the first
	case, and a \emph{counter-model} otherwise.
	This is an abuse of notation, and we need to know the exact nature of $P$ to
	know its precise meaning.
	In particular, we will study properties expressible on \emph{graph monadic second order
        logic} (MSO) over the signature
        $\{=, \to\}$, where
        $\to$ is a binary 
        relation such that $x \to y$ if and only if $y = f(x)$ or $y \in f(x)$
        (that is, $(x,y)\in E(\dyna{f})$).
        Our formulas express graphical properties of $\dyna{f}$,
        with logical operators ($\wedge$, $\vee$, $\neg$, $\implies$), and quantifications
        ($\exists,\forall$) on vertices (configurations) or sets of vertices
        (with the addition of $\in$ to the signature, for set membership).
	The \emph{quantifier rank} of a formula $\psi$ is its depth of quantifier
        nesting. %see for example \cite[Definition~3.8]{libkin04}.
	If $G$ and $G'$
	are two structures, we write $G \equiv_m G'$ when they satisfy the
	same formulas of quantifier rank $m$.
	We write $\psi \equiv \psi'$ when the two formulas $\psi$ and $\psi'$
	have the same models and the same counter-models.
	
        We will express general complexity lower bounds for non-trivial formulas,
        a notion which depends on the family of ANs or NANs under consideration.
	A formula $\psi$ is \emph{non-trivial} when it has infinitely 
        many models and infinitely many counter-models, among graphs of out-degree $1$.
        A formula $\psi$ is \emph{cw-non-trivial} when there exists $k$ such that $\psi$ has infinitely
        many models of cliquewidth at most $k$ and infinitely many counter-models of cliquewidth at most $k$
        (the definition of cliquewidth is recalled below).
        According to Remark~\ref{rem:allG}, these notions respectively correspond to deterministic
        ($\dyna{f}$ for ANs) and non-deterministic ($\dyna{f}$ for NANs) dynamics.
        When considering only graphs $\dyna{f}$ for $q$-uniform networks,
        we obtain the notions of \emph{$q$-non-trivial} and \emph{$q$-cw-non-trivial} formulas.
	%In the restriction where we only consider $q$-uniform ANs, $f$ is a \emph{model}
        %if and only if $f \vDash P$ and $f$ is $q$-uniform; $f$ is a \emph{counter-model}
        %if and only if $f \nvDash P$ and $f$ is $q$-uniform.
        %So if a formula $\psi$ is 
	%$\omega$-non-trivial in the general case,
        %it is not necessarily $\omega$-non-trivial in the restriction to $q$-uniform ANs.
	%%Je ne sais pas s'il ne faut pas enlever cette partie, pour définir directement au 
	%%début modele et contre-modele selon cette restriction

        In this article, the dynamical complexity of ANs and NANs is studied through
        an algorithmic point of view and the following families of decision problems,
        aimed at asking general properties on the behavior of a given network
        (the MSO formula $\psi$ is fixed in the problem definition).

        \medskip
        \noindent
	\decisionpbw{$\psi$-AN-dynamics}
        {circuit of an AN $f$ with $N$ configurations.}
        {does $\dyna{f} \vDash \psi$?}
        {.56}
	\decisionpbw{$\psi$-$q$-AN-dynamics}
        {circuit of a $q$-AN $f$ of size $n$.}
        {does $\dyna{f} \vDash \psi$?}
        {.42}
        
        \smallskip
        \noindent
	\decisionpbw{$\psi$-NAN-dynamics}
        {circuit of an NAN $f$ with $N$ configurations.}
        {does $\dyna{f} \vDash \psi$?}
        {.56}
	\decisionpbw{$\psi$-$q$-NAN-dynamics}
        {circuit of a $q$-NAN $f$ of size $n$.}
        {does $\dyna{f} \vDash \psi$?}
        {.42}
        \medskip

        \noindent
        Example formulas and associated \textbf{$\psi$-AN-dynamics} problems:
        \begin{itemize}[nosep]
          \item $\exists x, x\to x$ means that $f$ has a fixed point,
          %\item $\forall x, x\to x$ means that $f$ is the identity,
          \item $\exists x, \forall y, y\to x$ means that $f$ is a constant,
          \item $\forall x, \exists y, x\to y$ is trivially true,
          \item $\forall x,\forall y,\forall x',\forall y', (x\to y\wedge x'\to y'\wedge y=y')\implies x=x'$
            means that $f$ is injective,
          \item $\exists x,x\to x\wedge(\forall y,y\neq x\implies(\exists z,y\neq z\wedge y\to z\wedge z\to y))$
            requires $|V(\dyna{f})|$ to be odd,
            hence it is non-trivial in general,
            but trivial in the Boolean case ($2$-trivial).
        \end{itemize}
        For \textbf{$\psi$-NAN-dynamics} (non-deterministic) we have the following examples:
        \begin{itemize}[nosep]
          \item $\exists X, (\exists x \in X) \wedge (\forall x \in X, \exists y \in X: x \ne y \wedge x \to y)$
            means that $\dyna{f}$ has at least one non-trivial strongly connected component,
          \item $\forall x, \exists y, x\to y \wedge (\forall z, z\neq y \implies \neg(x\to z))$
            means that $\dyna{f}$ has out-degree $1$, which is trivial on deterministic dynamics
            (but it is cw-non-trivial).
        \end{itemize}
        Most formulas above do not use the second order quantifier,
        hence they are \emph{first order} (FO) formulas.
        It is known that properties such as connectivity, acyclicity, 2-colorability
        and planarity are expressible in MSO, but not in FO~\cite[Chapter~3]{libkin04}.

        Observe that the signature $\{=,\to\}$ only allows to express properties up to isomorphism,
        that is if $\dyna{f}$ and $\dyna{g}$ are isomorphic
        then $\dyna{f} \equiv_m \dyna{g}$ for all $m$.

        If $\psi$ is trivial (resp.~$q$-trivial), then \textbf{$\psi$-AN-dynamics}
        (resp.~\textbf{$\psi$-$q$-AN-dynamics}) has a finite list of
        positive instances or a finite list of negative instances,
        hence it is solvable in constant time.
        
        %Monadic Second Order logic, shortened by MSO, considers formulas with two kinds of
        %variables: vertices $(x_1, x_2, ...)$ and sets of vertices $(X_1, X_2, ...)$. It can be seen
        %as an extension of FO logic, in the sense that the vertices are the same. So we have the same
        %types of axiom on configurations as in the First-Order case, of the form $x \to y$ or $x =y$.
        %One kind of axiom that is added in MSO compared to FO is the inclusion written $x \in X$ where
        %$x$ is a configuration and $X$ a set of configurations. 
        %The notion of $\omega$-triviality is identical to the one of FO. 

	%%%%%%%%%%%%%%%%
	\subsection{Clique-decompositions and cliquewidth}
	
	For a finite set of $k$ colors $C = \{1,\dots,k\}$,
	a \emph{$k$-colored graph} $G = (V, E, C_G)$ 
        is a (directed) graph where each vertex is assigned one color by $C_G:V\to C$.
	%For a vertex $v \in G$, we denote $C_G(v)$ the color of $v$.
	The four following operations are used
        to construct $k$-colored graphs:
        \begin{enumerate}[nosep]
		\item $\constant i$ (arity 0): given a color $i \in C$, it is the graph 
		$G = (\{v\}, \emptyset, C_G)$ with $C_G(v) = i$ for some fresh vertex label $v$;
		\item constant$^\circ_i$ (arity 0): given a color $i \in C$, it is the graph 
		$G = (\{v\}, \{(v,v)\}, C_G)$ with $C_G(v) = i$ for some fresh vertex label $v$;
		\item $\recolor f$ (arity 1): given a \emph{recoloring} function $f : C \to C$ and a $k$-colored graph $H$, it is
		the graph $G = (V(H), E(H), C_G)$ such that $C_G = f \circ C_H$;
		\item $\join M$ (arity 2): given a set
		$M \subseteq C \times C \times \{R,L\}$
		and two $k$-colored graphs $H$ and $H'$, it
		is the graph $G = (V(H) \sqcup V(H'), E(H) \sqcup E(H') \sqcup E_R \sqcup E_L, C_G)$, where 
		$E_R = \{(u, v) \mid (i,j,R)\in M, u \in H, C_H(u) = i, v \in H',C_{H'}(v) = j\}$
		which are the arrows going from
		vertices of color $i$ in $H$ to vertices of color $j$ in $H'$ for triples with $R$ (right, from $H$ to $H'$),
		and symmetrically for
		$E_L = \{(u, v) \mid (i,j,L)\in M, u \in H', C_{H'}(u) = i, v \in H,C_H(v) = j\}$
		with $L$ (left, from $H'$ to $H$).
		Moreover, $C_G(v) = C_H(v)$ if $v \in H$ and
		$C_G(v) = C_{H'}(v)$ otherwise.
	\end{enumerate}

	Let $G$ be a $k$-colored graph. A \emph{clique-decomposition} $\mathcal{C}_G$ of $G$ is a binary tree 
	such that every node is labeled by one of these four operations,
        and the graph obtained at the root of $\mathcal{C}_G$ is $G$
        (the leaves of $\mathcal{C}_G$ are labeled by constant or constant$^\circ$, they are the vertices of $G$;
        nodes labeled by recolor have $1$ child; and nodes labeled by join have $2$ children).
        We say that $\mathcal{C}_G$ \emph{generates} the $k$-colored graph $G$, and that
	$\mathcal{C}_G$ is a clique-decomposition of a (non-colored) graph $G$ if there exists a coloring $C_G$
	of the graph $G$ such that $\mathcal{C}_G$ is a clique-decomposition of
	$(V(G), E(G), C_G)$.
        Observe that a clique-decomposition generates exactly one graph,
        hence we sometimes identify a clique-decomposition with the graph it generates %(colored or not) 
        (in particular, $|\mathcal{C}_G|$ will denote its number of constant nodes).
	%Saying that $\mathcal{C}_G$ is a clique-decomposition of a graph $G$ is the same 
	%thing as saying that a clique-decomposition $\mathcal{C}_G$ \emph{generates} a 
	%graph $G$ since one clique-decomposition generates exactly one graph. This is why 
	%with some abuse of notation talking about the clique-decomposition or the graph 
	%comes to the same.
	
	A \emph{marked clique-decomposition} is a clique-decomposition 
	with a unique \emph{marked} node where one child is replaced by a 
	node of arity $0$ labeled by $\square$. The marked node 
        acts as a join$_{\emptyset}$ (disjoint union). %as long as one of the child is $\square$.
	For a marked clique-decomposition $\mathcal{C}$ and a clique-decomposition 
	$\mathcal{C'}$, we denote $\mathcal{C} \oplus \mathcal{C'}$ the operation
	connecting them by replacing the $\square$ child of the marked node of $\mathcal{C}$
	with the root of $\mathcal{C'}$.
	We call this operation gluing $\mathcal{C}$ and $\mathcal{C'}$.
	The operation works the same when $\mathcal{C'}$ is also marked and in this case 
	$\mathcal{C} \oplus \mathcal{C'}$ is marked too, its marked node being the marked
	node of $\mathcal{C'}$.
	For two clique-decompositions $\mathcal{C}$ and $\mathcal{C'}$, we denote
	$\mathcal{C} \sqcup \mathcal{C'}$
	the clique-decomposition $\join{\emptyset}(\mathcal{C},\mathcal{C'})$,
        generating the disjoint union of the two graphs they respectively generate.
	
	The \emph{width} of a clique-decomposition is the number of colors used during 
	the generation of the colored graph (\emph{i.e.}~not necessarily only the colors present
	in the graph eventually generated at the root). The \emph{cliquewidth} of a graph $G$ is the minimal
	width of a clique-decomposition generating $G$.
	We say that a graph $G$ has a \emph{$k$-clique-decomposition}
	when $G$ can be generated by a clique-decomposition of width
	at most $k$.
	
        Let $\Gamma=\{\mathcal{C}_i\}_{i\in I}$ be a family of clique-decompositions
        ($\Gamma$ will always be a finite set),
	where only one $\mathcal{C}_j$ for some $j\in I$ is not marked;
        if $w=w_1\dots w_{\ell}$ is a nonempty word over alphabet $I$ where the only occurrence of $j$ is
	its last letter $w_{\ell}=j$, then define $\Delta^\Gamma(w)$
	by induction over the length of $w$ as follows:
	\begin{equation*}
		\Delta^\Gamma(w_1)=\mathcal{C}_{w_1},
		\hspace{1cm}
		\Delta^\Gamma(w_1\dots w_{n}) = \Delta^\Gamma(w_1\dots w_{n-1})\oplus \mathcal{C}_{w_{n}}.
	\end{equation*}
	
	See Figure~\ref{fig:tpack}. Observe that $\oplus$ is not commutative.
	In the case where $\Gamma$ is a set of graphs,
        by an abuse of notation,
	we consider $\Delta^\Gamma(w)$ to be the graph generated by 
	$\Delta^{\Gamma'}(w)$ where $\Gamma'$ is the set of clique-decompositions
	generating $\Gamma$.
	
	\begin{remark}
		\label{rem:different}
		Because of the recoloring functions, when taking two different nodes 
		in a clique-decomposition, the two clique-decompositions consisting of the 
		descendants of the two respective nodes might generate the same graph. Hence,
		in this article, when we say that two nodes $v$ and $v'$ are different in a 
		clique-decomposition or when writing $v \ne v'$, we mean that the two 
		clique-decompositions respectively consisting of the descendants of $v$ and 
		the descendants of $v'$ do not have the same set of constant nodes (hence
		the graph have different sets of vertices). 
		In particular, if $v'$ is itself a descendant of $v$, the clique-decomposition of the 
		descendants of $v$ contains at least one constant leaf which is not a descendant of $v'$.
	\end{remark}

	\subsection{Types}
	\label{ssec:types}
	
	Let $G$ be a $k$-colored graph, 
	$a_1,\dots,a_p$ be vertices of $G$ and $A_1,\dots,A_q$ be subsets of vertices of $G$.
	The \emph{quantifier rank $m$ type of $G,a_1,\dots,a_p,A_1,\dots,A_q$} 
	is the set of all MSO formulas with $p$ free point variables and $q$ free set variables and 
	quantifier rank at most $m$ which are satisfied by $G,a_1,\dots,a_p,A_1,\dots,A_q$.
	A set of formulas is a 
	\emph{realized $p,q$-type of quantifier rank $m$} if it is such a type.
	The following lemma is an essential key in 
	previous works and the present article:
	
	\begin{lemma}[{\cite[Proposition~7.5]{libkin04}}]\label{lemma:finitely-many-types}
		Given $p,q$ and $m$, there are finitely many realized $p,q$-types of quantifier rank $m$.
	\end{lemma}
	
	We denote $G,a_1,\dots,a_p,A_1,\dots,A_q \equiv_q H,b_1,\dots,b_p,B_1,\dots,B_q$ 
	when the corresponding types are the same.
	Our main focus will be on $0,0$-types which we simply 
        call \emph{types}.
	Let $T_m$ denote the (finite) set of realized types of quantifier rank $m$
        (it will always be clear from the context what is the fixed number $k$ of colors).
        We also attribute a type to each clique-decomposition (the type of the graph it generates),
        and to each node $v$ of a clique-decomposition
        (the type of the graph generated by the subtree whose root is $v$).
	
        %%%%%%%%%%%%%%%%
        \section{Previous works}
        \label{section:previous}

        The computational complexity of questions on the dynamics of automata networks
        has traditionally raised a great interest, as a mean to understand the possibilities and limits
        of algorithmic problem solving, in this model and multiple variants with restrictions
        on the kind of local functions (\emph{e.g.} Boolean, threshold, disjunctive),
        the architecture of the network, or with different update modes.
        Alon noticed in~\cite{a85} that it is $\NP$-hard to decide whether a given BAN has a fixed point.
        Flor\'een and Orponen then settled the complexity of multiple problems related to fixed points~\cite{fo89,o92}.
        Further developments~\cite{k08} include the study of limit cycles~\cite{bgmps21,sga22},
        update modes~\cite{psa16,npsv20,bgmps21,abgs23}
        and specific rules~\cite{gmpt18,gm20,gmrwt21}.

        In a recent series of two papers~\cite{ggpt21,gglgopt25},
        Rice-like complexity lower bounds have been established,
        encompassing at once many results from the literature.
        They state that non-trivial formulas yield algorithmically hard \textbf{$\psi$-dynamics} problems.
        The first article concentrates on the deterministic setting and FO formulas
        (using Hanf-Gaifman's locality),
        while the second article brings the result to MSO on non-deterministic automata networks
        (and introduces the notion of cw-non-trivial formula).

        \begin{customtheorem}{1.a for FO}[{\cite[Theorem~5.2]{ggpt21}}]
          \label{theorem:generalMainFO}
          For any non-trivial FO formula $\psi$,
          problem \textbf{$\psi$-AN-dynamics} is $\NP$-hard or $\coNP$-hard.
        \end{customtheorem}

        \begin{customtheorem}{1.c}[{\cite[Theorem~1]{gglgopt25}}] %gamard2023hardness
          \label{theorem:generalMainMSO}
          For any cw-non-trivial MSO formula $\psi$,
          problem \textbf{$\psi$-NAN-dynamics} is $\NP$-hard or $\coNP$-hard.
        \end{customtheorem}
        
        In the present work, we bring these results to $q$-uniform ANs and NANs, for MSO formulas
        (Theorem~\ref{theorem:hard}).
        Indeed, the reductions presented in the proofs of~\cite[Proposition~5.1.2 and Lemma~5.1.4]{ggpt21}
        and~\cite[Proposition~11 and Proposition~13]{gglgopt25}
        may require arbitrary alphabet sizes in the construction of instances of \textbf{$\psi$-AN-dynamics}
        and~\textbf{$\psi$-NAN-dynamics}, respectively.
        The proof of Theorem~\ref{theorem:hard} is based on the abstract pumping technique
        introduced in~\cite{ggpt21} and developed in~\cite{gglgopt25},
        which is reviewed in the rest of this section, including novel developments to establish a more precise construction
        for the purpose of further developments in the remaining sections of the present work.

        Let us start with the example of Figure~\ref{fig:AN}.
        It implements a reduction from \textbf{SAT}
        (here with $S=(x_1\vee x_2)\wedge(\neg x_1\vee\neg x_2)\wedge x_3$ on 3 Boolean variables),
        to prove the $\NP$-hardness of the fixed point existence problem ($\exists x:x\to x$).
        Indeed, for each valuation $v\in\{0,1\}^3$ of $S$,
        the pair of configurations $x,x'\in\{0,1\}^4$ with $x_{\{1,2,3\}}=x'_{\{1,2,3\}}=v$ creates either:
        \begin{itemize}[nosep]
          \item a path with a loop when $v\models S$, or
          \item a cycle of length $2$ when $v\not\models S$.
        \end{itemize}
        Observe that this same reduction also proves the $\coNP$-hardness
        of asking whether the dynamics is a union of limit cycles of length $2$,
        and of asking whether the dynamics is injective,
        both by reduction from \textbf{UNSAT}.
        The basic idea to reduce from \textbf{SAT} or \textbf{UNSAT}
        is to evaluate $S$ on the configuration,
        and produce (glue with $\oplus$) one of two graphs $G_0,G_1$
        (using a distinct subset of configurations for each valuation)
        according to whether the formula is satisfied or not.
        If the formula is satisfiable then $G_0$ appears at least once,
        otherwise there are only copies of $G_1$.
        The crucial part is then to obtain appropriate graphs $G_0$ and $G_1$
        for any non-trivial (or cw-non-trivial) MSO formula $\psi$,
        and this requires tools from finite model theory
        (giving additional parts $G_2,G_3$ to deal with).
        Analogously, the example of Figure~\ref{fig:NAN} is strongly connected
        if and only if $S$ is a tautology.

        Remark that $\psi$ is fixed in the problem definition,
        therefore $G_0,G_1$ are considered constant.
        Also, reductions produce circuits for the local functions, not the transition digraph itself.
        From~\cite{gglgopt25} %gamard2023hardness
        we first introduce the concept of saturating graph $G_0$ (Subsection~\ref{ss:satur}),
        then how $G_1$ has been obtained via the abstract pumping technique 
        that verify the requirements for the next subsection
        (Subsection~\ref{ssec:pump}) and finally 
        the metareduction (Subsection~\ref{ssec:sat}).
        We close this section with an overview of our contributions on strengthening this proof scheme
        (Subsection~\ref{ss:contrib}).

        %%%%
        \subsection{Saturating graph}
        \label{ss:satur}

        In~\cite{gglgopt25} it is shown that for all $m \in \N$, %gamard2023hardness
        there exists a \emph{saturating} graph $\Omega_m$ such that,
        for all MSO formulas $\psi$ of rank $m$,
        if a graph contains a copy of $\Omega_m$ then it is forced to be
        a model of $\psi$ (then we say that $\Omega_m$ is a \emph{sufficient subgraph} for $\psi$) or
        a counter-model of $\psi$ (then we say that $\Omega_m$ is a \emph{forbidden subgraph} for $\psi$).
        
        \begin{proposition}[{\cite[Proposition~7]{gglgopt25}}] %gamard2023hardness
        	\label{proposition:saturatingGraph}
        	Fix $m  \in \N$. There exists a graph $\Omega_m$ such that,
                for every MSO formula $\psi$ of rank $m$, either:
                \begin{enumerate}[nosep]
        		\item for every graph $G$ we have $G \sqcup \Omega_m \vDash \psi$, or
        		\item for every graph $G$ we have $G \sqcup \Omega_m \nvDash \psi$.
        	\end{enumerate}
        \end{proposition}

        The proof consists in considering $\Omega_m$ to be the disjoint union of sufficiently many copies
        of representatives from all equivalence classes of $\equiv_m$,
        so that $m$-round Ehrenfeucht–Fraïssé games cannot distinguish any additional copy ($G$ in the statement).
        It is immediate that the same proof applies when considering only
        graphs of out-degree $1$, because disjoint union preserves this property.

        \begin{corollary}
        	\label{corollary:saturatingGraph}
        	Fix $m  \in \N$. There exists a graph $\Omega_m$ of out-degree $1$ such that,
                for every MSO formula $\psi$ of rank $m$, either:
                \begin{enumerate}[nosep]
        		\item for every graph $G$ of out-degree $1$ we have $G \sqcup \Omega_m \vDash \psi$, or
        		\item for every graph $G$ of out-degree $1$ we have $G \sqcup \Omega_m \nvDash \psi$.
        	\end{enumerate}
        \end{corollary}
        
        It will be clear from the context whether we consider $\Omega_m$ with out-degree $1$:
        Proposition~\ref{proposition:saturatingGraph} for non-deterministic dynamics,
        and Corollary~\ref{corollary:saturatingGraph} for deterministic dynamics.

        Technically speaking, $\Omega_m$ is a fixed graph with some fixed cliquewidth $k'$.
        When taking the disjoint union with $G$ of cliquewidth at most $k$, the resulting graph $G\sqcup\Omega_m$
        has cliquewidth at most $\max\{k,k'\}$.
        In the sequel, the value of $k$ will be given by the hypothesis of cw-non-triviality
        on the fixed formula $\psi$ under study, and we will consider without loss of generality that $k\geq k'$
        so that $G\sqcup\Omega_m$ has cliquewidth at most $k$.
        This detail will be recalled when needed.
        
        %%%%
	\subsection{Pumping models of cw-non-trivial sentences}
	\label{ssec:pump}

	We pump in some clique-decompositions and construct a family of model clique-decompositions.
        Recall that since there is exactly one graph that corresponds to one clique-decomposition,
        the construction of model graphs follows.
	Let us sketch how three graphs are obtained in~\cite{gglgopt25}
        in order to construct an infinite family of models of $\psi$:
	
	\begin{proposition}[{\cite[Proposition~10]{gglgopt25}}]
		\label{proposition:generalConstruct}
		\label{proposition:generalConstruct4}
		Let $\psi$ be an MSO sentence and $k$ an integer.
		If $\psi$ has infinitely many models of cliquewidth at most $k$,
		then there exist clique-decompositions
		$\mathcal{C}_1,\mathcal{C}_2,\mathcal{C}_3$
		such that
		$
		\mathcal{C}_2 \oplus \mathcal{C}_1 \oplus \dots \oplus \mathcal{C}_1 \oplus \mathcal{C}_3
		$
		is a model of $\psi$, whatever the number of occurrences of $\mathcal{C}_1$.
                Furthermore, $\mathcal{C}_1$ contains at least one constant node (\emph{i.e.}, it is not the 
		trivial marked clique-decomposition with a unique node labeled by $\square$).
	\end{proposition}
	
	This proposition relies on the key concept of compositionality, stated as follows
        (recall that $T_m$ denotes the set of 
	realized types of quantifier rank $m$ for $k$-colored graphs):
	
	\begin{lemma}[{\cite[Lemma~9]{gglgopt25}}]
		\label{lem:compositionality-oplus}
		For every marked clique-decomposition $\mathcal{C}$ of width $k$ and every integer $m$, 
		there is a map $\Lambda_\mathcal{C}:T_m \to T_m$,
                such that for every clique-decomposition $\mathcal{C}'$ of type $t$,
                the type of $\mathcal{C} \oplus \mathcal{C}'$ is $\Lambda_\mathcal{C}(t)$.
                %such that for every clique-decomposition $\mathcal{C}'$, the type of the graph generated by 
		%$\mathcal{C} \oplus \mathcal{C}'$ is obtained from the type of the graph generated 
		%by $\mathcal{C}'$ by applying $\Lambda_\mathcal{C}$.
	\end{lemma}

        Lemma~\ref{lem:compositionality-oplus} means that when we have a marked clique-decomposition 
	$\mathcal{C}$ and two clique-decompositions $\mathcal{C}_1$ and $\mathcal{C}_2$
        of the same type, then 
	$\Lambda_\mathcal{C}(\mathcal{C}_1) = \Lambda_\mathcal{C}(\mathcal{C}_2)$. 
	In other words, $\mathcal{C} \oplus \mathcal{C}_1$ and
	$\mathcal{C} \oplus \mathcal{C}_2$ have the same type.
	The proof of Proposition~\ref{proposition:generalConstruct4} is
	illustrated on Figure~\ref{fig:pump}: any large enough clique-decomposition
        contains two vertices ($v$ and $v'$) of the same type (because $T_m$ is finite),
        consequently a piece of the clique-decomposition ($\mathcal{C}_1$)
        can be copied (pumped) any number of times while preserving the type of the obtained clique-decomposition
        (by Lemma~\ref{lem:compositionality-oplus}). By definition the type determines whether
        it is a model or a counter-model of $\psi$, hence Proposition~\ref{proposition:generalConstruct4} follows.
	
	\begin{figure}
		\newcommand{\myscale}{0.25}
		\begin{minipage}{0.38\linewidth}
			\centering
			\begin{tikzpicture}[scale=1.1*\myscale]
				% Drawing ranges from 0 to 16 in height
				\draw (0,0) -- +(3,3) node{$\bullet$} -- +(6,0) -- cycle;
				\draw (3,1.5) node{\tiny $\mathcal{C}_{w_{n}}$};
				\draw (0,3) -- +(3,3) node{$\bullet$} -- +(6,0) -- cycle;
				\draw (3,4.5) node{\tiny $\mathcal{C}_{w_{n-1}}$};
				\draw (0,6) -- +(3,3) -- +(6,0) -- cycle;
				\draw (3,7.5) node{\tiny $\mathcal{C}_{w_{n-2}}$};
				\draw (3,10.5) node{$\vdots$};
				\draw (0,12) -- +(3,3) node{$\bullet$} -- +(6,0) -- cycle;
				\draw (3,13.5) node{\tiny $\mathcal{C}_{w_2}$};
				\draw (0,15) -- +(3,3) -- +(6,0) -- cycle;
				\draw (3,16.5) node{\tiny $\mathcal{C}_{w_1}$};
			\end{tikzpicture}
			\caption{
				Illustration of $\Delta$. %$\tpack$
			} % Definition~\ref{dfn:tpack}
			\label{fig:tpack}
		\end{minipage}
		\begin{minipage}{0.58\linewidth}
			\centering
			\begin{tikzpicture}[scale=.55,baseline]
				\foreach \s [count=\i] in {0,1.5,3} {
					\coordinate (a\i) at ($(0,0)  +(0,-\s)$);
					\coordinate (b\i) at ($(2,3)  +(0,-\s)$);
					\coordinate (c\i) at ($(4,0)  +(0,-\s)$);
					\coordinate (d\i) at ($(3,0)  +(0,-\s)$);
					\coordinate (e\i) at ($(2,1.5)+(0,-\s)$);
					\coordinate (f\i) at ($(1,0)  +(0,-\s)$);
				}
				\foreach \i in {1,2} {
					\draw (a\i)--(b\i)--(c\i)--(d\i)--(e\i)--(f\i)--cycle;
				}
				\draw (a3)--(b3)--(c3)--cycle;
				\draw (b1)++(0,-.75) node{$\mathcal{C}_2$};
				\draw (b2) node{$\bullet$} node[right]{$v$}  ++(0,-.75) node{$\mathcal{C}_1$};
				\draw (b3) node{$\bullet$} node[right]{$v'$} ++(0,-1.5) node{$\mathcal{C}_3$};
				%\draw[decorate,decoration={brace,mirror}] (-.5,3) -- node[left]{$D_n$} (-.5,-3);
			\end{tikzpicture}
			\hspace{.5cm}
			$\to$
			\hspace{.5cm}
			\begin{tikzpicture}[scale=.55,baseline]
				\foreach \s [count=\i] in {0,1.5,...,6} {
					\coordinate (a\i) at ($(0,0)  +(0,-\s)$);
					\coordinate (b\i) at ($(2,3)  +(0,-\s)$);
					\coordinate (c\i) at ($(4,0)  +(0,-\s)$);
					\coordinate (d\i) at ($(3,0)  +(0,-\s)$);
					\coordinate (e\i) at ($(2,1.5)+(0,-\s)$);
					\coordinate (f\i) at ($(1,0)  +(0,-\s)$);
				}
				\foreach \i in {1,...,4} {
					\draw (a\i)--(b\i)--(c\i)--(d\i)--(e\i)--(f\i)--cycle;
				}
				\draw (a5)--(b5)--(c5)--cycle;
				\draw (b1)++(0,-.75) node{$\mathcal{C}_2$};
				\foreach \i in {2,...,4} {
					\draw (b\i)++(0,-.75) node{$\mathcal{C}_1$};
				}
				\draw (b5)++(0,-.75) node{$\mathcal{C}_3$};
			\end{tikzpicture}
			\caption{Proof of Proposition~\ref{proposition:generalConstruct4}.}
			\label{fig:pump}
		\end{minipage}
	\end{figure}
	
	\subsection{A reduction from SAT to {\bf $\psi$-NAN-dynamics}}
	\label{ssec:sat}
	
        In this section, we explain how Theorem~\ref{theorem:generalMainMSO}
        is obtained, and provide intermediate statements detailing a construction
        differing from the one presented in~\cite{gglgopt25}.
        It will be beneficial in order to explain the variants later,
        and we provide full proofs.

	To give an overview of this subsection,
	for a given MSO formula $\psi$ of rank $m$,
	whether $\Omega_m$ is sufficient (a model of $\psi$) or forbidden (a counter-model of $\psi$)
	indicates whether we respectively reduce from \textbf{SAT} ($\NP$-hardness)
	or \textbf{UNSAT} ($\coNP$-hardness).
	Given a propositional formula $S$ on $s$ variables,
	the configuration space $X$ is partitioned into $2^s$ subsets of configurations,
	so that $S$ is evaluated on each subset:
	\begin{itemize}[nosep]
		\item if $S$ is satisfied then those vertices create a copy of $\Omega_m$,
		\item if $S$ is falsified then those vertices create a neutral graph.
	\end{itemize}
	The neutral is intending to ``change nothing'', and is obtained by pumping (see 
	Section~\ref{ssec:pump}).
        Again, we consider clique-decompositions instead of the graphs they generate.

	%In all this section, for any biboundaried graph $G$, assume that $V(G)=\{0,\dots,|G|-1\}$
	%and for a node $u$ of $G$, write $G(u)$ the set $\{v:(u,v)\in E(G)\}$.
	%Thus, $P_1(G)$, $P_2(G)$ and $G(u)$ are all subsets of $\{0,\dots,|G|-1\}$.
	
	Recall that a clique-decomposition corresponds to a unique graph, therefore we abuse some notations.
	We first introduce a convenient way to deal with the assignments of $S$,
	then state our main result,
	and present our construction in a separate statement.
	
	\begin{definition}
		If $S$ is an instance of SAT with $s$ variables,
		then $\overline{S}$ is the word of length $2^s$ such that $\overline{S}_i$ is $1$ if $S(i)$ is false, and $0$ if $S(i)$ is true
		(viewing the binary expansion of $i$ as a Boolean assignment for $S$).
		And $\soo{S}$ is the word $\overline{S}$ where we insert an additional $1$
		in front of each symbol, hence of size $2^{s+1}$.
	\end{definition}
	
	For example, let $S=(x_1\vee\neg x_2)\wedge x_3$.
	Then $\overline S=11110010$, with $\overline S_7=0$ because the binary expansion of $7$ is $111$ and $(1\vee\neg 1)\wedge 1$ is true.
	And $\soo{S}=1111111110101110$.
	
	When $\mathcal{C}$ is a marked clique-decomposition and $\mathcal{C}'$
	is the clique-decomposition of a saturating graph,
	then we denote $\mathcal{C}\oplus\mathcal{C'}=\mathcal{C}\sqcup\mathcal{C'}$
	with the same marked node as $\mathcal{C}$.
	
	\begin{proposition}
		\label{pro:sat}
		Fix an MSO sentence $\psi$.
		Let $k$ be an integer and $\Gamma=\{\mathcal{C}_0, \mathcal{C}_1, \mathcal{C}_2, \mathcal{C}_3\}$
		four $k$-clique-decompositions satisfying the following conditions:
		\begin{enumerate}[label={(\roman*)},nosep]
			\item\label{itm:saturating} $\mathcal{C}_0$ is the clique-decomposition of a saturating graph;
			\item\label{itm:eqsize} $\mathcal{C}_0$ and $\mathcal{C}_1$ have the same number of constant nodes;
			\item\label{itm:marked} $\mathcal{C}_1$ and $\mathcal{C}_2$ are marked.
		\end{enumerate}
		Suppose that, for every word $w$ over alphabet $\{0,1\}$, we have $\Delta^\Gamma(2\cdot w \cdot 3)\models\psi$ if and only if $w$ contains letter $0$.
		Then {\bf $\psi$-NAN-dynamics} is $\NP$-hard. %, even if we restrict to inputs with bounded non-determinism.
	\end{proposition}
	
	%\begin{proposition}
	%  \label{pro:sat}
	%  Fix an MSO sentence $\phi$.
	%  Let $k$ be an integer and $\Gamma=\{G_0, G_1, G_2, G_3\}$ four $k$-graphs satisfying the following conditions:
	%  \begin{enumerate}[label={(\roman*)}]
		%  \item\label{itm:eqsz} $|G_0|=|G_1|$;
		%  \item\label{itm:cap} $P_1(G_0)\cap P_2(G_0)=P_1(G_1)\cap P_2(G_1)\neq V(G_1)$; and
		%  \item\label{itm:img} for every $p$ in $P_1(G_0)\cap P_2(G_0)$, we have $G_0(p)=G_1(p)$.
		%  \end{enumerate}
	%  Suppose that, for every word $w$ over alphabet $\{0,1\}$, we have $\gpack^\Gamma(2\cdot w \cdot 3)\models\phi$ if and only if $w$ contains letter $0$.
	%  Then \textsc{Succinct-}$\phi$ is \NP-hard. %, even if we restrict to inputs with bounded non-determinism.
	%\end{proposition}
	
        The reduction from a SAT instance $S$ must produce the circuit of a NAN whose transition digraph corresponds to
        a clique-decomposition obtained by pumping (Proposition~\ref{proposition:generalConstruct}),
        as illustrated on Figure~\ref{fig:sat}.
	Apart from some rework of the clique-decomposition to ensure convenient properties on the recolor nodes
	(stated in Lemma~\ref{lem:sat} and obtained in the proof of Proposition~\ref{pro:sat}),
	the construction presents no difficulty.
	It should also become clear why we interleave copies of $\mathcal{C}_1$.
	Given a $k$-clique-decomposition $\mathcal{C}$ with marked node $v$ and $C=\{1,\dots,k\}$,
	the composition of recolorings from $v$ to the root $r$ of $\mathcal{C}$ is simply the
	composition of functions $f:C\to C$ for all the recolor$_f$ nodes
	in the path from $v$ to $r$ (from the leaf to the root).
	From an initial color $c(v)$, it computes the color of node $v$ in the $k$-colored graph
	constructed by $\mathcal{C}$.
	A function $g:C\to C$ is \emph{idempotent} when $g^2=g$.
	
	\begin{figure}[t]
		\centering
		\begin{tikzpicture}[scale=.55,baseline]
			% pump C_1
			\foreach \s [count=\i] in {0,1.5,...,8} {
				\coordinate (a\i) at ($(0,0)  +(0,-\s)$);
				\coordinate (b\i) at ($(2,3)  +(0,-\s)$);
				\coordinate (c\i) at ($(4,0)  +(0,-\s)$);
				\coordinate (d\i) at ($(3,0)  +(0,-\s)$);
				\coordinate (e\i) at ($(2,1.5)+(0,-\s)$);
				\coordinate (f\i) at ($(1,0)  +(0,-\s)$);
			}
			\foreach \i in {1,2,3,5} {
				\draw (a\i)--(b\i)--(c\i)--(d\i)--(e\i)--(f\i)--cycle;
			}
			\draw (a6)--(b6)--(c6)--cycle;
			\draw (b1)++(0,-.75) node{$\mathcal{C}_2$};
			\foreach \i in {2,3,5} {
				\draw (b\i)++(0,-.75) node{$\mathcal{C}_1$};
			}
			\draw (b4)++(0,-.75) node{$\vdots$};
			\draw (b6)++(0,-.75) node{$\mathcal{C}_3$};
			\draw[decorate,decoration={brace}] (c2)++(.5,.5) -- node[right]{$13$ copies of $\mathcal{C}_1$} ++(0,-5.5);
			% union C_0 (désolé c'est pas aussi beau que le code du pompage...)
			\foreach \i in {1,2,3}
			\node[draw,rectangle,rounded corners] (j\i) at (-\i*2+1,7-\i*2) {join$_\emptyset$};
			\foreach [count=\i] \x/\y in {-1.7/2,-3.4/0,-6.6/0}{
				\coordinate (s\i) at (\x,\y);
				\draw (s\i)--++(1.5,-2)--++(-3,0)--cycle;
				\draw (s\i)++(0,-.75) node{$\mathcal{C}_0$};
			}
			\draw (s3)--(j3);
			\draw (s2)--(j3);
			\draw (j3)--(j2);
			\draw (s1)--(j2);
			\draw (j2)--(j1);
			\draw (b1)--(j1);
		\end{tikzpicture}
		\caption{
			Clique decomposition of $\Delta^\Gamma(2\cdot \soo{S} \cdot 3)$ in the statement of Lemma~\ref{lem:sat},
			for the word $\soo{S}=1111111110101110$ corresponding to the SAT instance $S=(x_1\vee\neg x_2)\wedge x_3$.
			Observe that disjoint unions for $\mathcal C_0$ employ the join$_\emptyset$.
		}
		\label{fig:sat}
	\end{figure}
	
	\begin{lemma}
		\label{lem:sat}
		Let $S$ be an instance of SAT with $s$ variables,
		$k$ an integer and $\Gamma=\{\mathcal{C}_0, \mathcal{C}_1, \mathcal{C}_2, \mathcal{C}_3\}$
		four $k$-clique-decompositions satisfying conditions~\ref{itm:saturating}--\ref{itm:marked}
		from Proposition~\ref{pro:sat}.
		Let $C=\{1,\dots,k\}$ and $g_1:C\to C$ be the composition of recolorings from the marked node
		of $\mathcal{C}_1$ to its root.
		If $g_1$ is idempotent,
		then there exists a NAN $f$ of size $n$ represented by circuit $D$
		that can be computed in polynomial time given $S$,
		such that
		$\Delta^\Gamma(2\cdot\soo{S}\cdot3)$
		is a clique-decomposition of
                $\dyna{f}$
		(the $\mathcal{C}_i$ are considered constant).
	\end{lemma}
	
	\begin{proof}
		Let $n_1$ be the number of constant nodes in $\mathcal{C}_0$ and $\mathcal{C}_1$
		which are equal by Condition~\ref{itm:eqsize},
		and $n_2$ (resp.~$n_3$) be the number of constant nodes in $\mathcal{C}_2$ (resp.~$\mathcal{C}_3$).
		We set $N=n_2+2^{s+1}\cdot n_1+n_3$, which is the number of constant nodes in
		$\Delta^\Gamma(2\cdot\soo{S}\cdot3)$.
		Recall that constant nodes define the vertices of $k$-colored graphs,
		hence we identify them in our explanations.
                In terms of the circuit $D$, the vertex labels of $\dyna{f}$
		are $\{0,\dots,n_2+2^{s+1}\cdot n_1+n_3-1\}$: see Figure~\ref{fig:satorga} for a picture of its organization.
		The initial segment corresponds to the (unique) copy of $\mathcal{C}_2$,
		followed by $2^{s+1}$ copies of $\mathcal{C}_0$ or $\mathcal{C}_1$,
		followed by the (unique) copy of $\mathcal{C}_3$.
		
		\begin{figure}[b]
			\centering
			\begin{tikzpicture}[xscale=.85,yscale=.7]
				\newcommand{\tichi}{0.1}
				\draw[->,>=latex] (0,0) -- (17,0);
				\draw (0,-\tichi) -- (0,+\tichi);
				\foreach \x in {0.2,0.4,...,16.8} {
					\draw (\x,0) -- (\x,\tichi);
				}
				\newcommand{\gr}[4]{
					\draw[decorate,decoration={brace}] (#1,#2) -- node[above]{#3} ++(#4,0);
					\draw[dashed] (#1,#2) -- (#1,0);
					\draw[dashed] (#1+#4,#2) -- (#1+#4,0);
				}
				\gr{0}{0.8}{$\mathcal{C}_2$}{1};
				\gr{1}{1.2}{$\mathcal{C}_1$}{3};
				\gr{4}{0.8}{$\mathcal{C}_0$ or $\mathcal{C}_1$}{3};
				\gr{7}{1.2}{$\mathcal{C}_1$}{3};
				\gr{10}{0.8}{$\mathcal{C}_0$ or $\mathcal{C}_1$}{3};
				\draw (14.4,0.8) node{$\dots$};
				\gr{15.8}{0.8}{$\mathcal{C}_3$}{1};
				
				\draw[<-] (0.1,-0.1) -- ++(0,-0.5) node[below]{$0$};
				\draw[<-] (1.1,-0.1) -- ++(0,-0.5) node[below]{$n_2$};
				\draw[<-] (4.1,-0.1) -- ++(0,-0.5) node[below]{$n_2+n_1$};
				\draw[<-] (7.1,-0.1) -- ++(0,-0.5) node[below]{$n_2+2\cdot n_1$};
				\draw[<-] (10.1,-0.1) -- ++(0,-0.5) node[below]{$n_2+3\cdot n_1$};
				\draw[<-] (15.9,-0.1) -- ++(0,-0.5) node[below]{$n_2+2^{s+1}\cdot n_1$};
			\end{tikzpicture}
			\caption{
				Proof of Lemma~\ref{lem:sat}, a representation of $\{0,\dots,n_2+2^{s+1}\cdot n_1+n_3-1\}$
				as the set of vertex labels of $\dyna{f}$ on $N$ configurations.
			}
			\label{fig:satorga}
		\end{figure}

        % (indentation décalée)

        In order to encode the NAN $f$ whose dynamics $\dyna{f}$ has clique-decomposition
	$\Delta^\Gamma(2\cdot\soo{S}\cdot3)$, the circuit $D$ needs to decide whether there is an edge
	from $v$ to $v'$.
	This depends on the set $M$ of their least common ancestor node join$_M$,
	and on their respective colors in the two children $k$-colored graphs.
	That is, the circuit needs to compute the result of recolor nodes upward.
	The idempotence condition makes this process finite and straightforward.
	
	Given $v$ and $v'$ two vertex labels in $\{0,\dots,N-1\}$,
	the circuit $D$ first computes the corresponding constant nodes in the clique-decomposition.
	It can belong to:
        \begin{itemize}[nosep]
		\item the copy of $\mathcal{C}_2$,
		\item one of the $2^{s+1}$ copies of $\mathcal{C}_0$ or $\mathcal{C}_1$,
		\item the copy of $\mathcal{C}_3$.
	\end{itemize}
	In order to compute the vertex label relative to a clique-decomposition from $\Gamma$
	(on a constant number of bits),
	and to keep track of the index among the $2^{s+1}$ copies of $\mathcal{C}_0$ or $\mathcal{C}_1$
	(on $s+2$ bits, to account for the unique copies of $\mathcal{C}_2$ and $\mathcal{C}_3$),
	we define respectively:
	\begin{align}
		\label{eq:relative}
		\delta(v)&=
		\begin{cases}
			v &\text{ if } v < n_2,\\
			(v-n_2) \mod n_1 &\text{ if } n_2 \leq v < n_2+2^{s+1}\cdot n_1,\\
			v-n_2-2^{s+1}\cdot n_1 &\text{ if } n_2+2^{s+1}\cdot n_1 \leq v;
		\end{cases}\\
		\label{eq:index}
		\gamma(v)&=
		\begin{cases}
			0 &\text{ if } v < n_2,\\
			1+\left\lfloor\frac{v-n_2}{n_1}\right\rfloor &\text{ if } n_2 \leq v < n_2+2^{s+1}\cdot n_1,\\
			2+2^{s+1} &\text{ if } n_2+2^{s+1}\cdot n_1 \leq v.
		\end{cases}
	\end{align}
	
	When $\gamma(v)\in\{2,4,6,\dots,2^{s+1}\}$, the circuit needs to evaluate $S$
	on the Boolean assignment given by the $s$ first bits of $\gamma(v)$, i.e., $\frac{\gamma(v)}{2}$,
	to decide whether $v$ belongs to a copy of $\mathcal{C}_0$ (true) or $\mathcal{C}_1$ (false).
	In all other cases it is straightforward to identify the clique-decomposition
	from $\Gamma$ to which $v$ belongs.
	
	In order to obtain the set $M$ of the node join$_M$ that is the least common ancestor of $v$ and $v'$,
	and to compute their respective colors in the two children $k$-colored graphs,
	we have the following cases.
        \begin{itemize}
		\item If $\gamma(v)=\gamma(v')$ then $v$ and $v'$ belong to the same copy of a clique-decomposition
		from $\Gamma$, and their adjacency is decided in constant time from their relative constant nodes 
		$\delta(v)$ and $\delta(v')$ in that clique-decomposition.
		\item Else, if $\gamma(v)\neq\gamma(v')$ and at least one of them belongs to a copy of $\mathcal{C}_0$,
		then there is no arc from $v$ to $v'$ because $M=\emptyset$ by the disjoint union of the saturating graph.
		\item Else, if $|\gamma(v)-\gamma(v')|\geq 3$, we present the case $\gamma(v)>\gamma(v')$
		(the other case is symmetric).
		Let $\mathcal{C}_x$ be the clique-decomposition of $v$ within $\Gamma$,
		and $\mathcal{C}_y$ be the clique-decomposition of $v'$ within $\Gamma$
		(we can assume $x,y\in\{1,2,3\}$).
		First, observe that the join$_M$ node belongs to the clique-decomposition of $v$:
		it is the join of $v$ and the marked node of $\mathcal{C}_x$.
		Consequently the color of $v$ in its child $k$-colored graph depends only on $\mathcal{C}_x$
		and is computed in constant time from $\delta(v)$.
		Second, observe that the path from $v'$ to this join$_M$ node contains at least one full traversal
		of a copy of $\mathcal{C}_1$ (here the interleaved copies of $\mathcal{C}_1$ are useful).
		Consequently, the color of $v'$ in its child $k$-colored graph is computed from its initial color $c(v')$
		as a constant node in $\mathcal{C}_y$,
		by applying: the finitely many recolorings in the path to the root of $\mathcal{C}_y$;
		then once $g_1$, which accounts for any number of copies of $\mathcal{C}_1$ by idempotency;
		and lastly the finitely many recolorings from the marked node of $\mathcal{C}_x$
		to the join$_M$ node.
		From the eventual colors of $v,v'$ and the set $M$, one decides whether the arc $(v,v')$ exists.
		\item Else, if $|\gamma(v)-\gamma(v')|\leq 2$, then there are finitely many cases to check
		(each requiring finite explorations from $\delta(v)$ and $\delta(v')$),
		depending on $\gamma(v),\gamma(v')$, their respective clique-decompositions within $\Gamma$,
		and whether they are separated by a copy of $\mathcal{C}_0$ or by a copy of $\mathcal{C}_1$
		(when $|\gamma(v)-\gamma(v')|=2$).
	\end{itemize}
	The algorithm described above can immediately be implemented as a circuit $D$
	produced in polynimial time, because the clique-decompositions $\mathcal{C}_i$ are considered
	constants, hence the only non-constant parts correspond to computing $\delta,\gamma$ and evaluating $S$
	(at most twice).
	This concludes the proof.
	\end{proof}
	
	We are now equipped to prove Proposition~\ref{pro:sat}, leading to Theorem~\ref{theorem:generalMainMSO}.
	The former bridges the gap to obtain the idempotency condition,
	and the latter assembles our Propositions from the previous sections.
	
	\begin{proof}[Proof of Proposition~\ref{pro:sat}]
		Given $\mathcal{C}_1$ from the statement, %of Proposition~\ref{pro:sat},
		consider $g_1:C\to C$ the composition of recolorings from the marked node of $\mathcal{C}_1$ to its root,
		with $C=\{1,\dots,k\}$.
		Since each element of a finite semigroup has an idempotent power,
		there is $p$ such that $g_1^p:C\to C$ is idempotent.
		Defining $\mathcal{C}'_1=\bigoplus_{i=1}^p \mathcal{C}_1$ and $\mathcal{C}'_0=\bigoplus_{i=1}^p \mathcal{C}_0$,
		the set $\Gamma'=\{\mathcal{C}'_0,\mathcal{C}'_1,\mathcal{C}_2,\mathcal{C}_3\}$
		fulfills all the conditions to apply Lemma~\ref{lem:sat}.
		It is now an immediate consequence of Lemma~\ref{lem:sat}:
		given an instance $S$ of SAT, compute in polynomial time the corresponding circuit %succinct graph representation
		$D$ of $f$ on $N$ configurations, then ask {\bf $\psi$-NAN-dynamics} on it. %$\mathcal{G}_{N,D}$
		% Moreover, the transition graphs of such automata networks are all of the form ${\Delta^\Gamma(w)}$ for a fixed set of finite graphs $\Gamma$ so they have a bounded degree, meaning that the constructed automata networks have bounded non-determinism.
	\end{proof}
	
	\begin{proof}[Proof of Theorem~\ref{theorem:generalMainMSO}]
		Let $\psi$ denote a cw-non-trivial MSO sentence with $m$ quantifiers.
		By Proposition~\ref{proposition:saturatingGraph}, the graph $\Omega_m$ is either forbidden or sufficient for $\psi$.
		Assume that $\Omega_m$ is sufficient for $\psi$ and prove that {\bf $\psi$-NAN-dynamics} is $\NP$-hard.
		(If it was forbidden, then consider $\neg\psi$ instead---this would yield that {\bf $\psi$-NAN-dynamics} is $\coNP$-hard.)
		Let $\mathcal{C}_0$ be a $k$-clique-decomposition of $\Omega_m$
                (this may require a larger $k$ than the value given by cw-non-triviality,
                but $\Omega_m$ is constant in our context,
                so we can just consider their maximum).
		By Proposition~\ref{proposition:generalConstruct4} applied to $\neg\psi$,
		there exist a triple of $k$-clique-decompositions $\Gamma=\{\mathcal{C}_1,\mathcal{C}_2,\mathcal{C}_3\}$
		such that $\Delta^\Gamma(2\cdot 1^n\cdot 3)$ is a model of $\neg\psi$ for every integer $n$.
		Let $\mathcal{C}'_1=\bigoplus_{i=1}^n \mathcal{C}_1$,
		where $n$ is the smallest integer such that the number of constant nodes in $\mathcal{C}'_1$
		is greater or equal to that of $\mathcal{C}_0$.
		Finally, let $\mathcal{C}'_0$ denote $\mathcal{C}_0$ with enough isolated vertices added so that
		its number of constant nodes equals that of $\mathcal{C}'_1$
		(adding an isolated vertex to a clique-decomposition corresponds to adding a constant node
		under a join$_\emptyset$ at the root).
		The quadruple $\Gamma'=\{\mathcal{C}'_0,\mathcal{C}'_1,\mathcal{C}_2,\mathcal{C}_3\}$ satisfies
		all the conditions of Proposition~\ref{pro:sat}.
		This concludes the proof of the theorem.
	\end{proof}
		
        Remark that in the proof of Theorem~\ref{theorem:generalMainMSO} above
        (in particular in the proof of Lemma~\ref{lem:sat} and its use in the proof of Proposition~\ref{pro:sat}),
        we have no asumption on the number $N$ of vertices (configurations)
        in the transition graph we produce.
        If $N$ is not a power of $q$, then it is impossible to get a $q$-uniform NAN.
        We solve this issue in this work.

        %%%%
        \subsection{Contributions}
        \label{ss:contrib}

        To prove Theorem~\ref{theorem:hard}, our contributions in this paper will be twofold.
        \begin{itemize}[nosep]
          \item A careful consideration in gluing (with $\oplus$) deterministic dynamics
            (graphs of out-degree $1$ are handled through Hanf's local lemma in~\cite{ggpt21})
            in Section~\ref{section:det}.
          \item The addition of arithmetical considerations in Section~\ref{section:lowerbounds},
            to restrict the constructions to $q$-uniform networks in Section~\ref{section:qunif}.
        \end{itemize}

	%%%%%%%%%%%%%%%%%%%%%%%%%%%%%%%%
	\section{Complexity lower bounds for deterministic networks}
	\label{section:det}
	
        In this section, we extend Theorem~\ref{theorem:generalMainFO} to MSO logics,
        or more accurately we transpose the proof technique of Theorem~\ref{theorem:generalMainMSO}
        from non-deterministic NANs to deterministic ANs (whose transition digraphs have out-degree $1$).
        Recall that the non-triviality of formulas is defined relative to graphs of out-degree $1$ only.
        We prove the following part of Theorem~\ref{theorem:hard}.
	
        \begin{customtheorem}{1.a}
		\label{theorem:detMainMSO}
		For any non-trivial MSO formula $\psi$,
		problem {\bf $\psi$-AN-dynamics} is $\NP$-hard or $\coNP$-hard.
	\end{customtheorem}
	
        We follow the proof structure presented in Section~\ref{section:previous},
        using the deterministic saturating graph from Corollary~\ref{corollary:saturatingGraph}.
        The main difficulty consists in transposing Proposition~\ref{proposition:generalConstruct4}
        in order to obtain graphs of out-degree $1$ all along the way
        (\emph{i.e.}, to ensure that gluings preserve this property),
        and meet %Conditions \emph{(i)--(iii)} of
        the required conditions for Proposition~\ref{pro:sat}.
        We combine these into the following proposition
        (out-degree $1$ graphs have cliquewidth at most $6$
        \footnote{Any graph of out-degree $1$ has treewidth at most 
        $2$ (its connected components are cycles with upward trees rooted on it),
        thus by~\cite{cr05} it has cliquewidth at most $6$.}).

	%We first show that it is possible to pump models and counter-models,
        %hence to obtain $G_0, G_1, G_2, G_3$ with the same property as in the general case, but more than that,
        %they will also verify for all $\ell \in \N$, that the graph $\Delta^{\Gamma}(2 \cdot w \cdot 3)$ is
	%of degree exactly $1$ for any word $w$ on alphabet $\{0, 1\}$. Hopefully, the reduction part is very
	%similar and the characteristics needed correspond well to the construction we will do here, 
	%so it gives a good idea of how the reduction works. We only prove the variant of 
	%Proposition~\ref{proposition:generalConstruct4} here, but to prove that the principle of reduction
	%in the non-deterministic case can also be adapted to the deterministic case, it will be shown in 
	%Section~\ref{section:logReduc}.
	
%	\begin{proposition}
%		\label{proposition:DetConstruct4}
%		Let $\psi$ be an MSO formula. If $\psi$ has infinitely many models of out-degree $1$,
%                then there exist $\tilde{\Gamma} = \{\tilde{G}_1, \tilde{G}_2, \tilde{G}_3\}$ three $3$-graphs such that
%                $\Delta^{\tilde{\Gamma}}(2 \cdot 1^{\ell} \cdot 3)\models\psi$ for all $\ell\in\N$,
%                and $P_1(\tilde{G_1})\cap P_2(\tilde{G_1})\neq V(\tilde{G_1})$.
%                Moreover, $\Delta^{\tilde{\Gamma}}(2 \cdot 1^{\ell} \cdot 3)$ has out-degree $1$ for all $\ell\in\N$.
%	\end{proposition}
\begin{proposition}
	\label{proposition:DetConstruct4}
	Let $\psi$ be an MSO sentence and $k$ an integer.
	If $\psi$ has infinitely many models of out-degree $1$,
	then there exist clique-decompositions
	$\mathcal{C}_1,\mathcal{C}_2,\mathcal{C}_3$
	such that
	$
	\mathcal{C}_2 \oplus \mathcal{C}_1 \oplus \dots \oplus \mathcal{C}_1 \oplus \mathcal{C}_3
	$
	is a model of $\psi$ of out-degree $1$, whatever the number of occurrences of $\mathcal{C}_1$.
	Furthermore, $\mathcal{C}_1$ contains at least one constant node.
\end{proposition}

	%To prove this proposition, as in non-deterministic case with 
	%Proposition~\ref{proposition:generalConstruct}, we prove the following theorem:
	%
	%\begin{proposition}
	%	\label{proposition:DetConstruct}
	%	Let $\psi$ denote an MSO sentence. If $\psi$ has infinitely many deterministic models of, 
	%	then there exist a triple of $3$-graphs $\Gamma = \{G_1, G_2, G_3\}$ such that
	%	$\Delta^{\Gamma}(2 \cdot 1^{\ell} \cdot 3)$ is a deterministic model of $\psi$ for every integer $\ell$.
	%\end{proposition}
	
        Our trick to prove Proposition~\ref{proposition:DetConstruct4} consists in introducing
        the property of having out-degree $1$ as an MSO formula.
        Let:
        \[
          %\chi = \forall x, \exists y, (x \to y) \wedge (\forall z, z \ne y \implies \neg (x \to z)),
          \chi = \left[\forall x, \exists y, (x \to y)\right] \wedge
          \left[\forall x, \forall y, \forall y', \neg(x\to y) \vee \neg(x\to y') \vee (y = y')\right],
        \]
        such that $G \models \chi$ if and only if the graph $G$ has out-degree exactly $1$
        (is a deterministic dynamics, or a functional graph).

	%The main difference with the non-deterministic proof is that instead of considering $\Sigma_{3, \psi}$-labeled
	%trees for tree-decompositions, we will consider $\Sigma_{3, \psi, \chi}$-labeled trees where $\psi$ and
	%$\chi$ are two MSO formulas, with a variant set of equivalence classes which will be of the form
	%$\Sigma_{3, \psi, \chi}$. The rest of the proof follows the same principle. 
	%
	%In the following, we note $\chi_1$ the following MSO formula:
	%$$\chi_1 = \forall x, \exists y, (x \to y) \wedge (\forall z, z \ne y \implies \neg (x \to z)),$$
	%such that a graph $G$ verifies $G \vDash \chi_1$ if and only if $G$ is a functional graph (of out-degree
	%exactly $1$).
	
%	Recall that $\Sigma_{k, \psi}$ denotes the set of equivalence classes of $\sim_{k, \psi}$ for $k$-graphs.
%        It is finite for any given $k$ and $\psi$, hence
%        $\Sigma_{k, \psi, \chi} = \Sigma_{k, \psi} \times \Sigma_{k, \chi}$
%        is finite as well.
%        In the rest of this section $k=3$, so we drop it from the notation.

	Recall that $T_m$ denotes the set of realized types of quantifier rank $m$, which is
	finite for any given $k$ and $m$. In particular, if 
	$m = \max(m', 2)$ where $m'$ is the quantifier rank of $\psi$ and $2$ the quantifier rank of
	$\chi$ and $k=6$, then $T_m$ is finite.
        In the following, when considering a $k$-clique-decomposition $\mathcal{C}$, 
        each node $v$ defines a maximal sub-clique-decomposition denoted
        $\mathcal{C}(v)$ (\emph{i.e.}, the clique-decomposition containing all the descendant of $v$) 
        that is rooted in $v$, and the type of $\mathcal{C}(v)$ is written $t(v)$.
	
%	Hence, for two nodes $v$ and $v'$ of a $T_q$-labeled tree, we have
%	$\mathcal{C}(v) = \mathcal{C}(v')$ if and only if 
%        $\mathcal{C}^\psi(v) = \mathcal{C}^\psi(v')$ and $\mathcal{C}^\chi(v) = \mathcal{C}^\chi(v')$.
%	We are now able to prove the proposition.

	\begin{proof} %[Proof of Proposition~\ref{proposition:DetConstruct}]
        	This proof brings the ideas of~\cite{gglgopt25}
                to the deterministic setting.
		Since $\psi$ has infinitely many deterministic models, 
%		we consider a family
%		$(M_i)_{i \in \N}$ of deterministic models and $(\mathcal{c_i)_{i \in \N}$ a sequence of their 
%                $3$-tree-decompositions, such that each $D_i$ has at least one path of $\mathcal{N}$-length $i$ 
%                (which is possible by splitting nodes of degree more than three in the tree).
%                Without loss of generality we assume that every $M_i$ has at least $3$ vertices,
%                and that all the bags of $D_i$ are of size $3$.
	and since there are finitely many types, there must be a type $t_0 \in T_m$ which contains 
	$\psi$ and $\chi$ and is realized by infinitely many graphs of cliquewidth at most $k$. Let 
	$\mathcal{C}$ be a clique-decomposition of cliquewidth at most $k$ of height at least 
	$|T_m|+1$, generating a graph of type $t_0$.
		
		%We view the tree-decompositions $D_i$ as $\Sigma_{\psi,\chi}$-labeled trees.
        %       Let $s=|\Sigma_{\psi, \chi}|$.
		Any path of length at least $|T_m|+1$ in a clique-decomposition contains
        at least two different nodes $v$ and $v'$ such that $t(v) = t(v')$.
        Let $v$ and $v'$ be two such nodes in a path of length at least $|T_m|+1$
        in $\mathcal{C}$. We assume without loss of generality that $v'$ is a
        descendant of $v$. Let $\mathcal{C}_3 = \mathcal{C}(v')$, $\mathcal{C}_1$ be
        $\mathcal{C}(v)$ where $\mathcal{C}(v')$ has been removed and $v'$ has been
        replaced by $\square$, and $\mathcal{C}_2$ be
        $\mathcal{C}$ where $\mathcal{C}(v)$ has been removed and $v$ has been
        replaced by $\square$.
        These definitions are such that 
        $\mathcal{C} = \mathcal{C}_2 \oplus \mathcal{C}_1 \oplus \mathcal{C}_3$. 
        Remark that $\mathcal{C}_1$ is not empty since $v \ne v'$ (Remark~\ref{rem:different}).
        
        Considering the map $\Lambda_{\mathcal{C}_1}$ from Lemma~\ref{lem:compositionality-oplus}, 
        since  the type of the graph generated by $\mathcal{C}_1 \oplus \mathcal{C}_3$ 
        is the same as the one of the graph generated by $\mathcal{C}_3$,
        we have
        $\Lambda_{\mathcal{C}_1}(t(v)) = t(v) = t(v') = \Lambda_{\mathcal{C}_1}(t(v'))$.
        So the type of the graph generated by 
        $\mathcal{C}_1 \oplus ... \oplus \mathcal{C}_ 1 \oplus \mathcal{C}_3$ is 
        $t(v)$, whatever the number of repetitions of $\mathcal{C}_1$. Finally, 
        since $\mathcal{C}= \mathcal{C}_2 \oplus \mathcal{C}_1 \oplus \mathcal{C}_3$ has type $t_0$, 
        we have $\Lambda_{\mathcal{C}_2}(t(v)) = t_0$ and the type of 
        $\mathcal{C}_2 \oplus \mathcal{C}_1 \oplus ... \oplus \mathcal{C}_1 \oplus \mathcal{C}_3$ 
        is always $t_0$, and the graph they are generating 
        are all models of both $\psi$ and $\chi$.
	\end{proof}

        Let $\Omega_m$ denote the deterministic saturating graph from Corollary~\ref{corollary:saturatingGraph},
        employed to reach the conditions of Proposition~\ref{pro:sat}.
	  \begin{proof}[Proof of Theorem~\ref{theorem:detMainMSO}]
	  	$\psi$ has an infinite number of deterministic models and deterministic counter-models
	  	of quantifier rank $q'$.
                Let $q = \max(q', 2)$, and suppose that 
	  	$\Omega_q \vDash \psi$. 
	        Since $\psi$ has infinitely many counter-models of out-degree $1$, let 
	        $\mathcal{C}_1, \mathcal{C}_2, \mathcal{C}_3$ be three clique-decompositions  
	        obtained by Proposition~\ref{proposition:DetConstruct4}, such that  $\mathcal{C}_1$
	 	contains at least one constant node and 
	  	$ \mathcal{C}_2 \oplus \mathcal{C}_1 \oplus \dots \oplus \mathcal{C}_1 \oplus \mathcal{C}_3$
is a model of $ \neg \psi$ of out-degree 1, whatever the number of occurrences of 
$\mathcal{C}_1$.

	  	Let $\mathcal{C}'_1=\bigoplus_{i=1}^n \mathcal{C}_1$,
	  	where $n$ is the smallest integer such that the number of constant nodes in $\mathcal{C}'_1$
	  	is greater or equal to that of $\mathcal{C}_0$.
	  	Finally, let $\mathcal{C}_0$ denote $\Omega_m$ with enough isolated vertices added so that
	  	its number of constant nodes equals that of $\mathcal{C}'_1$
	  	(adding an isolated vertex to a clique-decomposition corresponds to adding a constant node
	  	under a join$_\emptyset$ at the root).
	  	The quadruple $\Gamma'=\{\mathcal{C}_0,\mathcal{C}'_1,\mathcal{C}_2,\mathcal{C}_3\}$ satisfies
	  	all the conditions of Proposition~\ref{pro:sat}.
	  	In the case where $\Omega_m \nvDash \psi$, we have $\Omega_m \vDash \neg \psi$,
	  	hence the same proof applies to $\neg \psi$. 
	  	This concludes the proof of the theorem.
	  \end{proof}
	
	%%%%%%%%%%%%%%%%%%%%%%%%%%%%%%%%
        \section{Arithmetics for uniform networks}
        \label{section:lowerbounds}

        In this section we study the intersection of pairs of number sequences,
        to the purpose of constructing a metareduction proving Theorem~\ref{theorem:hard}
        for $q$-uniform automata networks (with a fixed alphabet of size $q$).
        We start by considering the restriction to Boolean alphabets,
        which is simpler. In Subsection~\ref{ss:boolean},
        we extract regularities in the intersection of the two sets of integers $\{ak+b \mid k\in \N_+\}$
        and $\{2^n \mid n \in \N\}$, respectively corresponding
        to the sizes of (counter-)model graphs produced by the pumping technique,
        and the sizes of admissible graphs as the dynamics of a Boolean automata network (on configurations $X=\bool^n$).
        In Subsection~\ref{ss:proof2}, we immediately sketch how to derive
        Theorem~\ref{theorem:hard} for $q=2$.
        In Subsection~\ref{ss:uniform}, we study the intersection of
        $\{ak+b \mid k\in \N_+\}$ and $\{q^n \mid n \in \N\}$,
	and again extract a geometric subsequence suitable for the reduction.
	Based on these arithmetical considerations,
        we will expose the proof of Theorem~\ref{theorem:hard} for $q$-uniform networks
        (for all $q\geq 2$) in Section~\ref{section:qunif}.
	%and generalize Theorem~\ref{theorem:hard} for any value of $q$
	%by using the result of Subsection~\ref{ss:uniform} to the case of $q$-uniform ANs.
	%In particular, the final construction of circuits for the \textbf{$\psi$-dynamics}
	%instance has simplified arithmetical considerations, thanks to the treatments from
	%Subsections~\ref{ss:boolean} and~\ref{ss:uniform}.
	%For the sake of being self-contained, it is nevertheless necessary to
	%reconsider the case disjunction on the structure of infinite (counter-)models
	%presented in Theorem~\ref{theorem:generalMainMSO}, with additionnal care on the
	%passage from general case to dynamics of ANs.

        %%%%%%%%%%%%%%%%
	\subsection{Geometric sequence for Boolean alphabets}
        \label{ss:boolean}
	
	Unless specified explicitly differently, we consider
	$(a,b) \in \N_+ \times \N$.
	Let:
	\[
	  \akb{a}{b}=\{ak+b \mid k\in \N_+\} \cap \{2^n \mid n \in \N\}
	\]
	be the integers which are both:
	\begin{itemize}[nosep]
	  \item the size of graphs obtained by pumping, \emph{i.e.}~from
            one base graph of size $b$ by gluing $k$ copies of graphs of size $a$,
	  \item the possible sizes of Boolean automata network dynamics,
	    \emph{i.e.}~of $|V(\dyna{f})|$.
	\end{itemize}
	We prove the following arithmetical theorem,
	aimed at being able to find suitable BAN sizes
	to design a polynomial time reduction to \textbf{$\psi$-dynamics}.
	That is, to exploit the pumping technique to produce automata networks with Boolean alphabets.
	The premise of the statement will derive from the non-triviality of $\psi$.
        In Subsection~\ref{ss:proof2}, we expose its
	consequences on Rice-like complexity lower bounds in the Boolean case.
	
	\begin{theorem}
		\label{theorem:interinf} 
		For all $(a,b) \in \N_+ \times \N$, if
		$\akb{a}{b}\neq\emptyset$ then it contains a geometric sequence of integers,
		and $|\akb{a}{b}|=\infty$. 
	\end{theorem}
	
	A \emph{solution} for $(a,b)$ is a pair $(K, N)\in\N_+ \times \N$ such that $aK
	+b = 2^N$. Remark that Theorem~\ref{theorem:interinf} straightforwardly holds
	when $b=0$, with solutions $(2^iK,N+i)$ for $i\in\N$, whenever one solution
	$(K,N)$ exists. We first discard another simple case.
	
	\begin{lemma} 
		\label{lemma:NotBodd} 
		If $b$ is odd and $a$ is even, then there
		is no solution, \emph{i.e.}~$\akb{a}{b}=\emptyset$. 
	\end{lemma}
	
	\begin{proof} 
		By contradiction, if $(K,N)$ is a solution, we can write $a =
		2a'$ and $b=2b'+1$, therefore $2a'K + 2b' + 1 = 2^N$. However $N\geq 1$, hence
		the left hand side is odd, whereas the right hand side is even, which is
		absurd. 
	\end{proof}
	
	The second lemma shows that we can restrict the study to a subset of $\N_+
	\times \N$.
	
	\begin{lemma} 
		\label{lemma:equivAB} 
		If Theorem~\ref{theorem:interinf} holds for
		any $(a,b)\in\N_+ \times \N$ with $a>b$ and $a$ is odd, then it holds for any
		$(a,b)\in\N_+ \times \N$. 
	\end{lemma}
	
	\begin{proof} 
		If $a \leq b$ we can write $b = am + b'$ with $m \in \N_+$ and
		$0\leq b' <a$ by Euclidean division. Observe that if $(K,N)$ is a solution for
		$(a,b)$, then $(K+m,N)$ is a solution for $(a,b')$, because
		$2^N=aK+b=a(K+m)+b'$. Conversely, if $(K,N)$ is a solution for $(a,b')$, then
		$(K-m,N)$ is a solution for $(a,b)$, provided that $K-m>0$. The geometric
		sequences contained in $\akb{a}{b}$ and $\akb{a}{b'}$ are identical, up to a
		shift of the first term. Consequently Theorem~\ref{theorem:interinf} holds for
		$(a,b)$ if and only if it holds for $(a,b')$, and we can restrict its study to
		the pairs $(a,b)$ with $a>b$.
		
		Assume $a>b$ and, without loss of generality, $b\neq 0$. 
		Let $\eta$ be the largest integer such that $2^{\eta}$ divides $a$, and
		$2^{\eta}$ divides $b$. Hence either $a'=\frac{a}{2^{\eta}}$ or
		$b'=\frac{b}{2^{\eta}}$ is odd (otherwise it would contradict the maximality
		of $\eta$). If $(a,b)$ has a solution $(K,N)$, then $2^{\eta}$ divides $2^N$.
		Moreover $a,b,K>0$, %it means that $2^N \geq a + b \geq 2 \times 2^{\eta}$
		therefore $N-\eta >0$, and $a'K + b' = 2^{N-\eta}$. As in the previous case,
		solving $\akb{a}{b}$ is similar to solving $\akb{a'}{b'}$, up to multiplying
		or dividing by $2^{\eta}$. If only $b'$ is odd, Lemma~\ref{lemma:NotBodd}
		concludes that Theorem~\ref{theorem:interinf} vacuously holds. Hence we can
		furthermore restrict the study of Theorem~\ref{theorem:interinf} to the pairs
		$(a,b)$ with $a>b$ and $a$ is odd. 
	\end{proof}
	
	The proof of Theorem~\ref{theorem:interinf} makes use of Fermat-Euler theorem:
	if $n$ and $a$ are coprime positive integers, then $a^{\varphi(n)} \equiv 1
	\mod n$, where $\varphi(n)$ is Euler's totient function.
	
	%\begin{theorem}[Euler's theorem]
	%  \label{theorem:euler}
	%  If $n$ and $a$ are coprime positive integers, and $\varphi(n)$ is Euler's
	%  totient function, then $$ a^{\varphi(n)} \equiv 1 \mod n.$$
	%\end{theorem}
	
	\begin{proof}[Proof of Theorem~\ref{theorem:interinf}] 
		Let
		$(a,b)\in\N_+\times\N$, with $a>b$, $a$ odd, and without loss of generality
		$b\neq 0$ hence $a>1$. Assume that $\akb{a}{b}\neq\emptyset$, \emph{i.e.}~there
		is a solution $(K,N)$ with $aK + b = 2^N$. Since $a$ is odd, $a$ and $2$ are
		coprime, and we can apply Fermat-Euler theorem: $2^{\varphi(a)} \equiv 1 \mod
		a$.
		
		Let $(u_\ell)_{\ell\in\N}$ be the geometric sequence %defined as $u_0=2^N$ and
		% $u_{\ell+1}=2^{\varphi(a)}u_{\ell}$, which gives
		$u_{\ell} = 2^{N + \ell \varphi(a)}$. We will now prove that $\{u_{\ell} \mid
		\ell \in \N \} \subseteq \akb{a}{b} = \{ak+b \mid k \in N_+\} \cap \{2^n | n
		\in \N\}$, %If we define the infinite geometric sequence $(u_{\ell})$ for
		% $\ell \in \N$
		%by $u_{\ell} = 2^{N + \ell \varphi(a)}$ (which is indeed geometric since we
		%can
		%also define it by induction as $u_0 = 2^N$ and $u_{\ell +1} =
		%2^{\varphi(a)}u_{\ell}$) then we can prove that $\{u_{\ell} | \ell \in \N \}
		%\subseteq \{ak+b | k \in N^*\} \cap \{2^n | n \in \N\}$,
		\emph{i.e.}~every $u_{\ell}$ can be written under the form $ak_{\ell} + b$ for
		$k_{\ell} \in \N_+$. Indeed:
		\begin{align*} 
			u_{\ell} &= 2^{N + \ell \varphi(a)} 
			= 2^{N} 2^{\ell \varphi(a)} \\ 
			&\equiv b 2^{\ell \varphi(a)} \mod a \\ 
			&\equiv b (2^{\varphi(a)})^{\ell} \mod a \\ 
			&\equiv b 1^{\ell} \mod a \\ 
			&\equiv b \mod a
		\end{align*}
		
		Hence for any $\ell$ there exists a $k_{\ell} \in \N$ such that $u_{\ell} =
		ak_{\ell} + b$. The sequence $(u_{\ell})_{\ell\in\N}$ is strictly increasing,
		so is $(k_{\ell})_{\ell\in\N}$. Furthermore $k_0=K>0$, so $k_\ell>0$ for all
		$\ell\in\N$. 
		
		We conclude that $\akb{a}{b}$ contains the geometric sequence $(u_\ell)_{\ell
			\in\N}$ for $a> b$ and $a$ odd, and Lemma~\ref{lemma:equivAB} allows to
		conclude for the general case $(a,b)\in\N_+\times\N$.
	\end{proof}
	
	%%%%%%%%%%%%%%%%
	\subsection{Proof of Theorem~\ref{theorem:hard} for $q=2$}
        \label{ss:proof2}
	
        In order to prove Theorem~\ref{theorem:hard} in the Boolean case $q=2$,
        the constructions are identical to the case of unrestricted alphabets,
        except that the graphs obtained by pumping and their number of copies in the metareduction
        are carefully chosen with the help of Theorem~\ref{theorem:interinf} above,
        in order to construct ANs or NANs whose sizes are powers of $2$.
        %Full details are given for any $q\geq 2$ in the proof of Theorem~\ref{theorem:unifMainMSO}
        %(Section~\ref{section:qunif}).

        \begin{customtheorem}{1 for $q=2$}
	  \label{theorem:2NPhard}
          For any $2$-non-trivial (resp.~$2$-cw-non-trivial) MSO formula $\psi$,
          problem {\bf $\psi$-2-AN-dynamics} (resp.~{\bf $\psi$-2-NAN-dynamics})
	  is $\NP$- or $\coNP$-hard.
	\end{customtheorem}

        The deterministic case is handled as in Section~\ref{section:det},
        to ensure that all the graphs have out-degree $1$.
        No other detail is needed, the main consideration here is on graph sizes.

        \begin{proof}
        Consider a $2$-non-trivial (resp.~$2$-cw-non-trivial) formula $\psi$.
        Without loss of generality, assume that the saturating graph
        $\Omega_m$ from Corollary~\ref{corollary:saturatingGraph} (resp.~Proposition~\ref{proposition:saturatingGraph})
        is a counter-model of $\psi$,
        and let us build a reduction from \textbf{UNSAT}
        (otherwise consider $\neg\psi$ instead of $\psi$,
        and a reduction from \textbf{SAT} instead of \textbf{UNSAT}).

        From Proposition~\ref{proposition:generalConstruct4}
        (resp.~Proposition~\ref{proposition:DetConstruct4})
        we obtain three clique-decompositions
        $\tilde{\Gamma}=\{\tilde{\mathcal{C}}_1,\tilde{\mathcal{C}}_2,\tilde{\mathcal{C}}_3\}$
        of cliquewidth at most $k$,
        but their construction does not ensure that the graph generated by
        $\Delta^{\tilde{\Gamma}}(2\cdot 1^\ell\cdot 3)$
        is the dynamics of a $2$-uniform automata network for any $\ell\in\N$.
        The only obstacle for this is the size of $\Delta^{\tilde{\Gamma}}(2\cdot 1^\ell\cdot 3)$,
        which is not guaranteed to be a power of $2$.
        From the construction, we only know that
        $\Delta^{\tilde{\Gamma}}(2\cdot 1\cdot 3)=
        \tilde{\mathcal{C}}_2\oplus\tilde{\mathcal{C}}_1\oplus\tilde{\mathcal{C}}_3$
        is a Boolean model of $\psi$ (from the $2$-non-triviality or $2$-cw-non-triviality).
        Hence, we deduce the existence of $(a,b)\in\N_+\times\N$ with:
        \begin{itemize}[nosep]
          \item $b=|\tilde{\mathcal{C}}_2\oplus\tilde{\mathcal{C}}_1\oplus\tilde{\mathcal{C}}_3|
            =|\tilde{\mathcal{C}}_2|+|\tilde{\mathcal{C}}_1|+|\tilde{\mathcal{C}}_3|$, %-|P_1(\tilde{G}_1)|-|P_1(\tilde{G}_3)|
          \item $a=|\tilde{\mathcal{C}}_1|$, %-|P_1(\tilde{G}_1)|
        \end{itemize}
        such that $|\Delta^{\tilde{\Gamma}}(2\cdot 1^\ell\cdot 3)|=a\ell+b$.
        When this size is a power of $2$, then $\Delta^{\tilde{\Gamma}}(2\cdot 1^\ell\cdot 3)$ is the dynamics of a BAN
        suitable for the reduction.
        We have the initial solution $(1,\log_2(a+b))$ for $\ell=1$ (where $a+b$ is a power of two),
        therefore by Theorem~\ref{theorem:interinf},
        in $\Delta^{\tilde{\Gamma}}(2\cdot 1^\ell\cdot 3)$
        we have %\footnote{Euler's totient function at $a\in\N_+$ is denoted $\varphi(a)$.}
        graphs of size $(a+b)2^{\ell'\varphi(a)}$ for all $\ell'\in\N$,
        precisely where $\ell=\frac{(a+b)2^{\ell'\varphi(a)}-b}{a}$ are integers
        (number of copies of $\tilde{\mathcal{C}}_1$, which is $1$ for $\ell'=0$).

        Before continuing, let us consider the fourth clique-decomposition.
        Let $\tilde{\mathcal{C}_0}=\mathcal{C}_m \sqcup \tilde{\mathcal{C}_1}$,
        where $\mathcal{C}_m$ is a clique-decomposition of $\Omega_m$.
        Let $\alpha$ be the least power of $2$ such that
        $\mathcal{C}_1=\bigoplus_{i=1}^{\alpha}\tilde{\mathcal{C}}_1$
        verifies $|\mathcal{C}_1|\geq|\tilde{\mathcal{C}}_0|$,
        and let $\mathcal{C}_0$ be the disjoint union of $\tilde{\mathcal{C}}_0$
        and $|\mathcal{C}_1|-|\tilde{\mathcal{C}}_0|$ constant$^\circ$ nodes,
        so that $|\mathcal{C}_0|=|\mathcal{C}_1|$.
        Finally, let $\mathcal{C}_2=\tilde{\mathcal{C}}_2$ and $\mathcal{C}_3=\tilde{\mathcal{C}}_3$.
        The clique-decompositions $\Gamma=\{\mathcal{C}_0,\mathcal{C}_1,\mathcal{C}_2,\mathcal{C}_3\}$
        of cliquewidth at most $k$ verify
        Conditions \emph{(i)--(iii)} of Proposition~\ref{pro:sat}.
        Furthermore, $\mathcal{C}_1$ is made of $\alpha$ copies of $\tilde{\mathcal{C}}_1$.

        Given a propositional formula $S$ on $s$ variables,
        we need at least $2^s$ copies of $\mathcal{C}_0$ or $\mathcal{C}_1$ in order to implement the reduction from \textbf{UNSAT},
        where for each valuation the circuit of the AN (resp.~NAN) produces
        a copy of $\mathcal{C}_0$ if $S$ is satisfied, and a copy of $\mathcal{C}_1$ otherwise
        (see Subsection~\ref{ssec:sat}).
        Recall that $\mathcal{C}_1$ is made of $\alpha$ copies of $\tilde{\mathcal{C}}_1$
        and $|\mathcal{C}_0|=|\mathcal{C}_1|$, that is, for $|w|=2^s$
        we have $|\Delta^\Gamma(2\cdot w\cdot 3)|=a\alpha 2^s+b$.
        The goal is now to find a suitable padding by $L(s)$ copies of $\tilde{\mathcal{C}}_1$,
        \emph{i.e.}~such that:
        \[
          |\Delta^{\Gamma'}(2\cdot w \cdot 4^{L(s)}\cdot 3)|=a(\alpha 2^s+L(s))+b
        \]
        is a power of $2$, with $|w|=2^s$, $\mathcal{C}_4=\tilde{\mathcal{C}}_1$
        and $\Gamma'=\{\mathcal{C}_0,\mathcal{C}_1,\mathcal{C}_2,\mathcal{C}_3,\mathcal{C}_4\}$.
        Then the reduction is as usual: the graph $\Delta^{\Gamma'}(2\cdot w \cdot 4^{L(s)}\cdot 3)$
        is a model of $\psi$ if and only if $w$ does not contain letter $0$ 
        (in this case it has only copies of $\tilde{\mathcal{C}}_1$).
        Remember that the only non-constant part is $S$ (and $s$).

        We want to find $\ell'\in\N$ such that:
        \[
          \alpha 2^s+L(s) = \frac{(a+b)2^{\ell'\varphi(a)}-b}{a}
          \iff
          L(s) = 2^{\ell'\varphi(a)}+\frac{b(2^{\ell'\varphi(a)}-1)}{a}-\alpha 2^s.
        \]
        A solution with $L(s)\geq 0$ is $\ell'=s+\log_2(\alpha)$, because the middle term is
        positive\footnote{$\ell'=\left\lceil\frac{s+\log_2(\alpha)}{\varphi(a)}\right\rceil$
        would be a smaller valid solution, but ceilings are not handy.
        For the same reason, $\alpha$ has been chosen to be a power of $2$.}.
        Therefore, we can build a Boolean AN (resp.~Boolean NAN) with
        \begin{align*}
          n=\log_2(a+b)+\log_2(\alpha)\varphi(a)+s\varphi(a) &\quad\text{automata, and}\\
          \alpha 2^s+L(s)=\frac{(a+b)\alpha^{\varphi(a)}2^{s\varphi(a)}-b}{a} &\quad\text{copies of $\tilde{\mathcal{C}}_1$}
        \end{align*}
        (some copies of $\tilde{\mathcal{C}}_1$ form copies of $\mathcal{C}_1$, or are replaced by a copy of
        $\mathcal{C}_0$ when $S$ is satisfied).
        Recall that $a$, $b$, $\varphi(a)$ and $\alpha$ are constants depending only on $\psi$.

        The circuit first identifies whether the input configuration $x\in\bool^n$ belongs to:
        \begin{itemize}[nosep]
          \item the copy of $\mathcal{C}_2$: from $0$ to $|\mathcal{C}_2|-1$,
          \item one of the $2^s$ copies of $\mathcal{C}_0$ or $\mathcal{C}_1$ depending on $S$:
            from $|\mathcal{C}_2|$ to $|\mathcal{C}_2|+2^s|\mathcal{C}_1|-1$,
            where $\beta=\lfloor\frac{x-|\mathcal{C}_2|}{|\mathcal{C}_1|}\rfloor$
            indicates the valuation from $0$ to $2^s-1$, and
            $x-|\mathcal{C}_2|-\beta|\mathcal{C}_1|$ indicates the position within a copy of $\mathcal{C}_0$ or $\mathcal{C}_1$,
          \item the copy of $\mathcal{C}_3$ at the end of the gluing:
            from $|\mathcal{C}_2|+2^s|\mathcal{C}_1|$ to $|\mathcal{C}_2|+2^s|\mathcal{C}_1|+|\mathcal{C}_3|-1$,
          \item one of the $L(s)$ copies of $\mathcal{C}_4=\tilde{\mathcal{C}}_1$:
            from $|\mathcal{C}_2|+2^s|\mathcal{C}_1|+|\mathcal{C}_3|$ to $2^n-1$.
        \end{itemize}
        Then it outputs the image of $x$ (deterministic case),
        or determines whether the second input configuration $y$ is an out-neighbor of $x$ (non-deterministic case).
        \end{proof}
	
        %%%%%%%%%%%%%%%%
	\subsection{Geometric sequence for $q$-uniform alphabets}
        \label{ss:uniform}

        Let us consider an arbitrary integer $q \geq 2$.
	We adapt the developments of Subsection~\ref{ss:boolean},
        in order to produce $q$-uniform networks (of size $q^n$)
        via the pumping technique.
        We denote $\N_{++}=\N\setminus\{0,1\}$, so that $q\in\N_{++}$.
	Unless specified explicitly differently, we consider
	$(a,b) \in \N_+ \times \N$.
	Let
        \[
          \akbq{a}{b}{q}=\{ak+b \mid k\in \N\} \cap \{q^n \mid n \in \N\}.
        \]
        Our goal is to prove the following arithmetical result.
	
	\begin{theorem} 
		\label{theorem:interinfQ} 
		For all $(a,b,q) \in \N_+ \times \N \times \N_{++}$, if 
		$\akbq{a}{b}{q}\neq\emptyset$ and there exist 
                $(\kappa,\mu) \in \N \times \N_+$ such that 
		$a \kappa + b = b q^{\mu}$, then $\akbq{a}{b}{q}$ contains a
		geometric sequence of integers and $|\akbq{a}{b}{q}|=\infty$.
                Otherwise, if for 
                any $(\kappa, \mu) \in \N \times \N_+$, $a\kappa + b \ne bq^{\mu}$, 
		then $|\akbq{a}{b}{q}| \le 1$.
	\end{theorem}

	A \emph{solution} for $(a,b,q)$ is a pair $(K, N)\in\N^2$
	such that $aK +b = q^N$.
	Remark that Theorem~\ref{theorem:interinfQ} straightforwardly
	holds when $b=0$, with solutions $(q^iK,N+i)$ for $i\in\N$, whenever one
	solution $(K,N)$ exists (and this solution is necessarily with $K \ne 0$). 

        Theorem~\ref{theorem:interinfQ} is slightly more involved than
        Theorem~\ref{theorem:interinf} for the Boolean case ($q=2$),
	because we need to consider the case $k=0$ in the definition of $\akbq{a}{b}{q}$,
	for the purpose of the proof of Theorem~\ref{theorem:hard}
        in Section~\ref{section:qunif}.
	For example, $a=2$, $b=4$, $q=2$ verifies
	$|\akbq{a}{b}{q}|=\infty$, and our base case in the proof of Theorem~\ref{theorem:hard}
	will be the solution $K=0$, $N=2$.
	Consequently, we must also deal with examples such as
	$a=4$, $b=2$, $q=2$ where $|\akbq{a}{b}{q}|=1$
	(the unique solution is $K=0$, $N=1$).

        Remark that $\kappa = 0$ implies that there exists $\mu \in \N_+$ such that 
	$b=bq^{\mu}$, but since $q \ge 2$ we have $b=0$.
	In the following we will only consider the case $\kappa \in \N_+$.

	We distinguish two parts in the statement of Theorem~\ref{theorem:interinfQ},
        the first being the case with conclusion $|\akbq{a}{b}{q}| = \infty$,
        and the second with conclusion $|\akbq{a}{b}{q}| \le 1$.
        After proving Theorem~\ref{theorem:interinfQ},
        we will formulate another condition equivalent to the existence of $(\mu,\kappa)\in\N\times\N$
        such $a\kappa+b=bq^\mu$, adapted to the base case of the pumping technique.

	%In the next lemma, we prove why we need in the first part of
	%Theorem~\ref{theorem:interinfQ} the restriction to $(a,b,q)$ such that there
	%exist $\mu, \kappa \in \N_+ \times  \N_+$ verifying $b q^{\mu} = a \kappa + b$,
        %by showing that otherwise we have at most one solution
        %(second part of Theorem~\ref{theorem:interinfQ}).
	Let us start by proving the second part of Theorem~\ref{theorem:interinfQ},
	emphasizing that its first part has necessary conditions.
	%Recall that we can restrict the study to $\kappa \ne 0$.
	
	\begin{lemma}
		\label{lemma:conditionAB} 
		If for any $(\kappa,\mu) \in \N_+ \times \N_+$,
		$a\kappa + b \ne bq^{\mu}$, then $|\akbq{a}{b}{q}| \le 1$. 
	\end{lemma}
	
	\begin{proof} 
		For the sake of a contradiction, assume that we have at least two
		solutions for $(a,b,q)$, denoted $(K,N)$ and $(K',N')$.
                We consider without loss of generality that $K' > K$ and $N' > N$.
		By definition of solution we have $q^{N'} = aK' + b$. Moreover, we have
		$q^{N'} = q^N q^{N'-N} = (aK+ b)q^{N'-N}$.
                It follows that:
                \begin{align*}
                  aK' + b &= (aK+b)q^{N'-N}\\
                  \iff a(K' - Kq^{N'-N}) + b &= bq^{N'-N}.
                \end{align*}
		We necessarily have $K' - Kq^{N'-N} > 0$ since $bq^{N'-N} > b$.
		This contradicts the hypothesis on $a,b,q$, hence $|\akbq{a}{b}{q}| \le 1$.
	\end{proof}

        To study the first part of Theorem~\ref{theorem:interinfQ},
	we discard another simple case.
		
	\begin{lemma} 
		\label{lemma:NotBdivisibleQ} 
		If $b$ is not divisible by $q$ and
		$a$ is divisible by $q$, then 
		either there is a unique solution $(K, N)\in\N^2$ with $N=0$,
		%\emph{i.e.}~$|\akbq{a}{b}{q}|=1$,
		%and for any 
		%$\mu, \kappa \in \N_+ \times \N_+$ we have $b q^{\mu} \ne a \kappa + b$;
                or there is no solution, \emph{i.e.}~$\akbq{a}{b}{q}=\emptyset$.
	\end{lemma}
	
	\begin{proof}
		If $(K,N)$ is a solution, we can write
		$a = qa'$ and $b=qb'+q'$, with $0 < q' < q$ not divisible by $q$, therefore 
		$qa'K + qb' + q' = q^N$.
		By contradiction, if $N\geq 1$, the left hand side is not divisible by
		$q$, whereas the right hand side is divisible by $q$, which is absurd.
		Hence there is no solution with $N\geq 1$.
		Otherwise, for $N=0$, since we discarded the case $b=0$, necessarily 
		$K=0$, $b'=0$ and $q'=1$, so there is a unique solution $(0,0)$.
		%it is clear that there are not any 
		%$\mu, \kappa \in \N_+\times \N_+$
		%such that $q^{\mu} = qa'\kappa + 1$.
	\end{proof}
	
	The first part of Theorem~\ref{theorem:interinfQ} can be restricted
	to a subset of triplets from $\N_+ \times \N \times \N_{++}$,
	strengthening the hypothesis for the rest of the proof.
	
	\begin{lemma} 
		\label{lemma:equivABQ} 
		If the first part of Theorem~\ref{theorem:interinfQ} holds
		for any $(a,b,q)\in\N_+ \times \N \times \N_{++}$ with $a>b$ and $a$ not
		divisible by $q$, then it holds for any $(a,b,q)\in\N_+ \times \N \times
                \N_{++}$. 
	\end{lemma}
	
	\begin{proof}
		If $a \leq b$ we can write $b = am + b'$ with $m \in \N_+$ and
		$0\leq b' <a$ by Euclidean division. Observe that if $(K,N)$ is a solution for
		$(a,b, q)$, then $(K+m,N)$ is a solution for $(a,b', q)$, because
		$q^N=aK+b=a(K+m)+b'$. Conversely, if $(K,N)$ is a solution for $(a,b', q)$,
		then $(K-m,N)$ is a solution for $(a,b, q)$, provided that $K-m>0$. The
		geometric sequences contained in $\akbq{a}{b}{q}$ and $\akbq{a}{b'}{q}$ are
		identical, up to a shift of the first term. Consequently the first part of
		Theorem~\ref{theorem:interinfQ} holds for $(a,b, q)$ if and only if it holds
		for $(a,b', q)$, and we can restrict its study to the triplets with $a>b$.
		
		Assume $a>b$ and, without loss of generality, $b\neq 0$. Let $\eta$ be the
		largest integer such that $q^{\eta}$ divides $a$, and $q^{\eta}$ divides $b$.
		Hence either $a'=\frac{a}{q^{\eta}}$ or $b'=\frac{b}{q^{\eta}}$ is not
		divisible by $q$ (otherwise it would contradict the maximality of $\eta$). If
		$(a,b, q)$ has a solution $(K,N)$, then $q^{\eta}$ divides $q^N$. 
		So $N-\eta \ge 0$, and $a'K + b' = q^{N-\eta}$. As in the
		previous case, solving $\akbq{a}{b}{q}$ is similar to solving
		$\akbq{a'}{b'}{q}$, up to multiplying or dividing by $q^{\eta}$. If only $b'$
		is not divisible by $q$, Lemma~\ref{lemma:NotBdivisibleQ} concludes that
		Theorem~\ref{theorem:interinfQ} vacuously holds (otherwise, we would be 
		in the second part of Theorem~\ref{theorem:interinfQ}, which is a contradiction
		with the hypothesis). Hence we can furthermore
		restrict the study of the first part of Theorem~\ref{theorem:interinfQ} 
		to the triplets with $a>b$ and $a$ not divisible by $q$. 
	\end{proof}

        Before proving the first part of Theorem~\ref{theorem:interinfQ},
        we express a simple proposition of arithmetics,
        aimed at showing that our pumping hits desirable sizes.

	\begin{proposition} 
		\label{proposition:recB} 
		For all $(a,b,q,n) \in \N_+ \times \N_+ \times \N_{++} \times \N_+$, 
		if $b q^n \equiv b \mod a$ then for any $m \in
		\N_+$, we have $b q^{mn} \equiv b \mod a$. 
	\end{proposition}
	
	\begin{proof} 
		We prove the lemma by induction on $m$.
		The base case $m =1$ is immediate by hypothesis.
		For the induction, let $m \ge 2$.
                We have:
		\begin{align*} 
			b q^{mn} &= b q^{(m-1)n} q^n \\ 
                        &\equiv b q^n \mod a \quad\textnormal{(by induction hypothesis)}
			\\ 
                        &\equiv b \mod a\\[-3em]
		\end{align*}
	\end{proof}

        We are ready to prove the first part of Theorem~\ref{theorem:interinfQ}.
	Recall that we consider $\kappa\neq 0$.

	\begin{lemma} 
		\label{lemma:interinfQ} 
		For all $(a,b,q) \in \N_+ \times \N
		\times \N_{++}$, if $\akbq{a}{b}{q}\neq\emptyset$ and there exist $(\kappa,\mu) \in
		\N_+ \times \N_+$ such that $a \kappa + b = b q^{\mu}$, then $\akbq{a}{b}{q}$ 
		contains a geometric sequence of integers and $|\akbq{a}{b}{q}|=\infty$. 
	\end{lemma}
	
	\begin{proof} 
		By Lemma~\ref{lemma:equivABQ}, it is sufficient to prove the claim for
		$a>b$ and $a$ not divisible by $q$.
		Assume that $\akbq{a}{b}{q}\neq\emptyset$, \emph{i.e.}~there is a solution
		$(K,N)$ with $aK + b = q^N$.
		Assume furthermore that we have $(\mu, \kappa) \in \N_+ \times \N_+$ 
		such that $a \kappa + b = bq^{\mu}$.
		
		Let $(u_\ell)_{\ell\in\N}$ be the geometric sequence defined as
		$u_{\ell} = q^{N + \ell \mu}$.
		We will now prove that
		$\{u_{\ell} \mid \ell \in \N \}
		\subseteq
		\akbq{a}{b}{q} = \{ak+b \mid k \in N\} \cap \{q^n | n \in \N\}$,
		\emph{i.e.}~every $u_{\ell}$ can be written under the form $ak_{\ell} + b$ for some $k_{\ell}
		\in \N$.  Indeed:
		\begin{align*} 
			u_{\ell} &= q^{N} q^{\ell \mu} \\
			&\equiv b q^{\ell \mu} \mod a \\ 
                        &\equiv b \mod a \quad\textnormal{(by
                                Proposition~\ref{proposition:recB})}
		\end{align*}
		Hence for any $\ell$ there exists $k_{\ell} \in \N$ such that $u_{\ell} =
		ak_{\ell} + b$. The sequence $(u_{\ell})_{\ell\in\N}$ is strictly increasing,
		so is $(k_{\ell})_{\ell\in\N}$. 
%		Furthermore $k_0=K>0$, so $k_\ell>0$ for all
%		$\ell\in\N$.
		%We conclude that $\akbq{a}{b}{q}$ contains the geometric sequence
		%$(u_\ell)_{\ell \in\N}$ for $a>b$ and $a$ not divisible by $q$, and
		%Lemma~\ref{lemma:equivABQ} allows to conclude for the general case
		%$(a,b)\in\N_+\times\N$. 
	\end{proof}
	
	\begin{proof}[Proof of Theorem~\ref{theorem:interinfQ}] 
		By
		Lemmas~\ref{lemma:conditionAB} and~\ref{lemma:interinfQ}
                (recall that $\kappa=0$ implies $b=0$).
	\end{proof}

        We now study in more details when the condition of Theorem~\ref{theorem:interinfQ} 
        holds.  Indeed, we will apply it to models in the next section, so we want to find a condition
        easier to verify. In the rest of this section we prove such an equivalent condition.
        
        For any $(a,q) \in \N_+\times\N_{++}$, define the \emph{coprime power of $q$ for $a$}
        as the smallest integer $\eta$ such that when writing $a = a' \gcd(a, q^{\eta})$,
        $a'$ is coprime with $q$.
	
	\begin{proposition}
		\label{proposition:existCoprimePower}
		For any $(a,q) \in \N_+\times\N_{++}$, the coprime power of $q$ for $a$ exists.
	\end{proposition}
	
	\begin{proof}
		We write the \emph{prime factorization} of $a$ and $q$ as $a = \prod_{p \in
	\mathcal{P}}p^{v_p(a)}$ and $q = \prod_{p \in \mathcal{P}}p^{v_p(q)}$, where
$\mathcal{P}$ is the set of prime numbers and $v_i(j)$ is the biggest integer
such that $i^{v_i(j)}$ divides $j$. 
%		Let $d \in \N_+$ be defined such that for any $p \in \mathcal{P}$, $v_p(d) =
%0$ if $v_p(q) =0$ and $v_p(d) = v_p(a)$ otherwise ($v_p(q) \ne 0$).
%
%		By construction, $d$ divides $a$.  Moreover, since all the prime numbers
%appearing in the prime factorization of $d$ are also appearing in the prime
%factorization of $q$, there exists a smallest integer $\eta \in \N_+$ such that
%$d$ divides $q^{\eta}$.
%
%		Let $a', q' \in \N_+$ such that $a = a'd$ and $q^{\eta}= q'd$. By construction
%of $d$ with only prime factors of $q$, $a'$ is coprime with $q$.
%		
%		In fact $\eta$ is the smallest integer such that for all $p \in \mathcal{P}$
%if $v_p(q) \ne 0$ then $v_p(q^{\eta}) \ge v_p(a)$. So in fact $d = \gcd(a,
%q^{\eta})$.
	Let:
        \[
          \eta = \max \{ \left\lceil \frac{v_p(a)}{v_p(q)} \right\rceil \mid v_p(q) \ne 0 \}.
        \]
	Observe that $\eta$ is well defined %$\{ \lceil \frac{v_p(a)}{v_p(q)} \rceil \mid v_p(q) \ne 0 \} \ne \emptyset$ 
	since $q \ge 2$. We prove that $\eta$ is the coprime power of $q$ for $a$. 
	It holds that:
        \[
          \gcd(a, q^{\eta}) = \prod_{p \in \mathcal{P}}p^{\min(v_p(a), \eta v_p(q))} 
          = \prod_{p \in \mathcal{P}, v_p(q) \ne 0}p^{v_p(a)}.
        \]
	Moreover, defining $a' = \prod_{p \in \mathcal{P}, v_p(q) = 0}p^{v_p(a)}$, 
	which is by construction coprime with $q$, we have $a = a' \gcd(a, q^{\eta})$.
	
	Finally, we prove that $\eta$ is the smallest integer with such a property. For the 
	sake of a contradiction, assume that there exists $\mu < \eta$ with the same 
	property. Consequently, there exists at least one prime number $p \in \mathcal{P}$ such that
	$v_p(q) \ne 0$ and  $\min(v_p(a), \mu v_p(q)) < \min(v_p(a), \eta v_p(q)) = v_p(a)$. 
	Hence, $a = a'' p \gcd(a, q^{\mu})$ for some $a'' \in \N_+$ and $a''p$ is not 
	coprime with $q$, which is a contradiction.
	\end{proof}
	
	In the next lemma, we show a sufficient condition on $(a,b,q)$ such that it
        verifies the conditions on the first part of Theorem~\ref{theorem:interinfQ}.
        As developed above we assume that $b>0$ (hence $a>1$), $\kappa>0$,
        and also by Lemma~\ref{lemma:equivABQ} that $a > b$.

	\begin{lemma} 
		\label{lemma:suffCond} 
                If $(a,b,q) \in \N_+ \times \N_+ \times \N_{++}$ with $a>1$ are
		such that there is a solution $(K,N)$ to $\akbq{a}{b}{q}$,
                and the coprime power $\eta$ of $q$ for $a$ verifies $\eta \le N$,
                then there exist 
                $(\mu, \kappa) \in \N_+ \times \N_+$
		such that  $bq^{\mu} = a \kappa + b$. 
	\end{lemma}
	
	The proof of Lemma~\ref{lemma:suffCond} uses Fermat-Euler theorem,
        which we recall:
        if $n$ and $a$ are coprime positive integers, then $a^{\varphi(n)} \equiv 1 \mod n$,
        where $\varphi(n)$ is Euler's totient function.

	\begin{proof} 
		Assume that there is a solution $(K,N)$ such that $aK +b = q^N$,
		and that the coprime power of $q$ for $a$, denoted $\eta$,
		verifies $\eta \le N$. We write $a = a' \gcd(a, q^{\eta})$ and 
		$q^{\eta} = q' \gcd(a, q^{\eta})$. 
		Since $\eta \le N$, we have $q^N = q^{N-\eta}q'\gcd(a, q^{\eta})$.
		
		If $a' = 1$, it means that we have $q^N = q^{N-\eta}q'a$. Hence $aK + b =
		q^{N-\eta}q'a$ and $b = (q^{N-\eta}q'-K)a$, in other words $b \equiv 0 \mod
		a$. So for any $\mu \in \N_+$ we have $bq^{\mu} \equiv  b \mod a$. In
		particular $\mu = 1$ is sufficient.
		
		If $a' > 1$, then by Fermat-Euler theorem ($a'$ and $q$ are coprime by
		definition of coprime power) we have 
		$q^{\varphi(a')} \equiv 1 \mod a'$,
                and denote $r\in\N$ such that $q^{\varphi(a')} = r a' +1$.
                We also use the fact that $b = q^N - aK = \gcd(a,
		q^{\eta})(q^{N-\eta}q' - a'K)$:
		\begin{align*} 
                  b q^{\varphi(a')} &= b (ra' +1)\\
                  &=\gcd(a,q^\eta)(q^{N-\eta}q'-a'K)(ra'+1)\\
                  &= \big(r(q^{N-\eta}q'-a'K)\big) \big(a' \gcd(a,q^\eta)\big) + \gcd(a,q^\eta)(q^{N-\eta}q'-a'K)\\
                  &= \big(r(q^{N-\eta}q'-a'K)\big) a + b\\
                  &\equiv b \mod a
		\end{align*} 
		Therefore, with $\mu = \varphi(a')$, there exists $\kappa \in \N_+$ such that 
		$b q^{\mu} = a\kappa +b$.
	\end{proof}

        The following remark elucidates why the case $q=2$ (and any prime $q$) offers
        simpler considerations on the arithmetics of pumping.

	\begin{remark} 
		If $q$ is prime, then we can assume that $a$ is coprime with $q$
                (otherwise we employ Lemma~\ref{lemma:equivABQ}, because $a$ is divisible by $q$).
                As a consequence, $a' = a$ and $\eta = 0$
		in Lemma ~\ref{lemma:suffCond}, so the
		sufficient condition is verified for any $b \in \N_+$,
                and by Theorem~\ref{theorem:interinfQ}
                there is a geometric sequence of integers in $\akbq{a}{b}{q}$.
	\end{remark}
	
%	\begin{remark} 
%		In fact, the sufficient condition of Lemma~\ref{lemma:suffCond}
%		is not on the existence of a minimal $\eta$ but on the verification of the
%		inequality $\eta \le N$. Indeed for any $x,y \in \N_+^2$ there always exists
%		an
%		$n \in \N_+$ such that if we write $x = x' \gcd(x, y^n)$ then $x'$ is coprime
%		with $y$ (the proof can be made with prime factorizations). 
%	\end{remark}
	
	The following lemma shows that the sufficient condition of
	Lemma~\ref{lemma:suffCond} is necessary. 
	
	\begin{lemma} 
		\label{lemma:necessCond}
                If $(a,b,q) \in \N_+ \times \N_+ \times \N_{++}$ with $a>1$ are
                such that there is a solution $(K, N)$ to $\akbq{a}{b}{q}$,
                and the coprime power $\eta$ of $q$ for $a$ verifies $\eta > N$,
                then for any $(\mu,\kappa) \in \N_+ \times \N_+$, $bq^{\mu} \ne a\kappa + b$. 	
	\end{lemma}
	
	\begin{proof} 
		Assume that there is a solution $(K,N)$ such that $aK +b = q^N$,
		and that the coprime power of $q$ for $a$, denoted $\eta$,
		verifies $\eta>N$. We write $a = a' \gcd(a, q^{\eta})$ and 
		$q^{\eta} = q' \gcd(a, q^{\eta})$. 
		
                For the sake of a contradiction, suppose that there exist
                $(\mu,\kappa) \in \N_+\times\N_+$ such that $b q^{\mu} = a \kappa + b$.
                By Proposition~\ref{proposition:recB}, we know that $bq^{\ell \mu} \equiv b \mod a$
                for all $\ell \in \N_+$. Consider an $\ell \in \N_+$ such that $\ell \mu \ge \eta$,
                and denote $\kappa_{\ell} \in \N_+$ such that $b q^{\ell \mu} = a\kappa_{\ell} +b$.
		
		We have $bq^{\ell \mu - \eta}  q^{\eta} = a\kappa_{\ell} +b$. Given
		that $\gcd(a, q^{\eta})$ divides both $a$ and $q^{\eta}$, it also divides 
		$bq^{\ell \mu - \eta}  q^{\eta} - a\kappa_{\ell} $ which is $b$.
		Consequently, $\gcd(a, q^{\eta})$ divides $aK +b$, which is $q^N$.
                Hence $q^{\eta} = q^{\eta - N} q''\gcd(a, q^{\eta}),$ with $q''$ such that
                $q^N = q''\gcd(a,q^{\eta})$ (and $q' = q^{\eta - N} q''$).
                It contradicts the minimality of $\eta$,
                therefore such $(\mu,\kappa)\in\N_+\times\N_+$ does not exist. 
	\end{proof}

        Combining Lemmas~\ref{lemma:suffCond} and~\ref{lemma:necessCond}, we obtain an equivalent condition.

        \begin{theorem}
                \label{theorem:equivCond}
                For $(a,b,q) \in \N_+ \times \N_+ \times \N_{++}$ with $a>1$,
                there exist $(\mu,\kappa) \in \N_+ \times \N_+$ such that $bq^{\mu} = a\kappa + b$
                if and only if
                there is a solution $(K, N)$ to $\akbq{a}{b}{q}$ and
                the coprime power $\eta$ of $q$ for $a$ verifies $\eta \leq N$.
        \end{theorem}
        
        Note that from the developments above, for any $(a,b,q) \in \N_+ \times \N_+ \times \N_{++}$
        with $a>1$ and $\eta$ the coprime power of $q$ for $a$,
        it is impossible to have two solutions $(K,N)$ and $(K',N')$
        such that $\eta > N$ and $\eta \leq N'$.
        The equivalent condition of Theorem~\ref{theorem:equivCond},
        in order to apply Theorem~\ref{theorem:interinfQ},
        essentially states that if we manage to find a solution to $\akbq{a}{b}{q}$
        from a large enough model (graph) compared to the pumpable part (with $N$ big compared to $a$),
        then the pumping does intersect powers of $q$ regularly, according to a geometric sequence.
	
        %%%%%%%%%%%%%%%%
	\section{Complexity lower bounds for $q$-uniform networks}
        \label{section:qunif}
        
        In this part we prove our main results,
        that the Rice-like complexity lower bound holds on $q$-uniform automata networks,
        both deterministic and non-deterministic, for any alphabet size $q\geq 2$.
        %In this part we will prove the following variant of Theorem~\ref{theorem:generalMainMSO}:
        
        \begin{customtheorem}{1.b,d}
        	\label{theorem:unifMainMSO}
                Let $q\geq 2$.
                For any $q$-non-trivial (resp.~$q$-cw-non-trivial) MSO formula $\psi$,
                problem {\bf $\psi$-$q$-AN-dynamics} (resp.~{\bf $\psi$-$q$-NAN-dynamics})
                is $\NP$- or $\coNP$-hard.
        \end{customtheorem}

        The deterministic case is handled as in Section~\ref{section:det},
        to ensure that all the graphs have out-degree $1$.
        No other detail in needed, the main consideration here is on graph sizes.

        Again, the major difficulty is in the adaptation of Proposition~\ref{proposition:generalConstruct4}
        to $q$-uniform networks, \emph{i.e.}~to ensure that the pumping gives graphs whose sizes are powers of $q$.
        To this purpose, we will employ Theorem~\ref{theorem:interinfQ}
        and the equivalent formulation of its condition in Theorem~\ref{theorem:equivCond}.
        This latter is expressed in terms of the coprime power of $q$ for $a$,
        but $a$ (the size of the pumped graph) is not known at this stage.
        In order to overcome this issue,
        we first fix adequate representatives for each MSO type of graph,
        hence with a finite number of sizes.
        Afterwards, we will be able to get a value upper bounding the coprime power requirement
        of Theorem~\ref{theorem:equivCond}, to unlock the arithmetics with Theorem~\ref{theorem:interinfQ}.

        %In this section, even if it is not written explicitly for every proposition or theorem, every proof will be done in a 
        %way that if we need to preserve the determinism it can be preserved. For that, the ideas will be the same as in 
        %Section~\ref{section:det}.

        %%%%%%%%%%%%%%%%
        \subsection{Type representatives}
		
	Recall that $T_m$ is the set of realized types of quantifier rank $m$ and it is finite.
	In the following, we denote $t(\mathcal{C})$ the type of the graph generated by 
	$\mathcal{C}$, the latter being written $G_{\mathcal{C}}$. Moreover, for a node
	$v$ in $\mathcal{C}$ we still denote $\mathcal{C}(v)$ the clique-decomposition
        composed by $v$ and all its descendant (the subtree whose root is $v$),
        and $t(v)=t(\mathcal{C}(v))$.
	
	Let $T'_m \subseteq T_m$ be the set of realized types of quantifier rank $m$ such 
	that, for each type $t \in T'_m$, there exists a representative clique-decomposition $\tau$ such that:
	\begin{itemize}[nosep]
		\item $t(\tau) = t$,
                \item there is a node $v\in \tau$ which is not the root such that $t(\tau(v)) = t$.
	\end{itemize}
        Let $s' = |T'_m|$, and denote $\{\tau_1,\dots,\tau_{s'}\}$ those representatives.
        The first condition helps identifying representatives,
        and the second condition will correspond to the base case for pumping models
        (by gluing copies of $\tau(v)$ inside $\tau$), which will be given by our hypothesis
        on the non-triviality or the cw-non-triviality of the formula $\psi$.
        For all $i\in\int{s'}$,
        %let $\tau_i$ be a $k$-clique-decomposition of $\sigma_i$ and
        let $\tau'_i$ be $\tau_i$ where $\tau_i(u)$ has been removed and $u$ has been
        replaced by $\square$.
        %and such that $\tau'_i$ is generating $\sigma'_i$. These
        %definitions are such that $\tau_i = \tau'_i \oplus \tau(u)$ and that the generated graphs
        %are coherent with that.
        Hence $\tau_i = \tau'_i \oplus \tau(u)$.

        Given that these representatives are fixed and form a finite set,
        we will be able to upper bound the coprime power
        from Section~\ref{section:lowerbounds} for the arithmetics of pumping $q$-uniform dynamics,
        independently of the initial model.
        Indeed, in the next construction we will substitute the pumped part (called $\tilde{\mathcal{C}}_1$ so far)
        with its representative among $\{\tau'_i\}_{i\in\int{s'}}$,
        because its type will belong to $T'_m$ by hypothesis.
        
        %%%%%%%%%%%%%%%%
        \subsection{Unbounded pumping}
        
        In order to meet the condition of Theorem~\ref{theorem:equivCond}
        and pump on graph sizes which are powers of $q$,
        we extend Proposition~\ref{proposition:generalConstruct4} by proving that
        it is possible to pump on models of arbitrary size.
  
        \begin{proposition}
        	\label{proposition:unifConstruct4}
                Let $x$ be an arbitrary integer,
        	$\psi$ an MSO formula and an integer $k$.
                If $\psi$ has infinitely many $q$-uniform 
        	models of cliquewidth at most $k$,
                then there exist $\tilde{\Gamma} = \{\tilde{\mathcal{C}}_1,\tilde{\mathcal{C}}_2,\tilde{\mathcal{C}}_3\}$
                three clique-decomposition of width at most $k$ such that
                $\Delta^{\tilde{\Gamma}}(2 \cdot 1^{\ell} \cdot 3)\models\psi$
                for all $\ell\in\N$.
                Moreover, $|\tilde{\mathcal{C}}_2\oplus\tilde{\mathcal{C}}_3|=q^N$ for $N$ verifying $x \leq N$,
                and $\tilde{\mathcal{C}}_1=\tau'_i$ for $i\in\int{s'}$.
        \end{proposition}

        We will apply Proposition~\ref{proposition:unifConstruct4} for some $x$ related
        to the coprime powers of representatives $\{\tau'_i\}_{i\in\int{s'}}$,
        but the following proof does not rely on any distinctive feature of coprime powers,
        therefore we state it for an arbitrary value $x$.
        
        \begin{proof}
          As in the proof of Proposition~\ref{proposition:DetConstruct4}, since $\psi$ has infinitely many $q$-uniform models, 
          	and since there are finitely many types, there must be a type $t_0 \in T_m$ which contains 
          	$\psi$ and is realized by infinitely many $q$-uniform graphs of cliquewidth at most $k$. Let 
          	$\mathcal{C}$ be a clique-decomposition of cliquewidth at most $k$ of height at least 
                $|T_m|+1$, such that $t(\mathcal{C})=t_0$ and $|\mathcal{C}=q^N$ for some
                %generating a graph $G$ of type $t_0$ such that $|G| = q^N$ for some 
                $N \in \N$. Without loss of generality we assume that $\mathcal{C}$ has at least $q^x$ constant nodes
          	(otherwise, we would just have to consider a bigger $q$-uniform model).
          	
          	%We view the tree-decompositions $D_i$ as $\Sigma_{\psi,\chi}$-labeled trees.
          	%       Let $s=|\Sigma_{\psi, \chi}|$.
          	Any path of length at least $|T_m|+1$ in a clique-decomposition contains
          	at least two different nodes $v$ and $v'$ such that $t(\mathcal{C}(v)) = t(\mathcal{C}(v'))$.
          	Let $v$ and $v'$ be two such nodes in a path of length at least $|T_m|+1$
          	in $\mathcal{C}$. We assume without loss of generality that $v'$ is a
          	descendant of $v$. 
          	Because of the characteristics of $\mathcal{C}(v)$, there is a representative 
          	$\tau_i \in T_m'$ such that $t(\mathcal{C}(v)) = t(\tau_i)$,
          	we denote it $t$.
          	
          	Let $\tilde{\mathcal{C}_3} = \mathcal{C}(v)$, $\tilde{\mathcal{C}_1} = \tau_i'$, 
          	and $\tilde{\mathcal{C}_2}$ be $\mathcal{C}$ where $\mathcal{C}(v)$ has been removed 
          	and $v$ has been replaced by $\square$. These definitions are such that 
          	$\mathcal{C} = \tilde{\mathcal{C}_2} \oplus \tilde{\mathcal{C}_3}$,
          	see Figure~\ref{figure:qunifPump}.
                Remark that $\mathcal{C}_1$ is not empty since $v \ne v'$ (Remark~\ref{rem:different}).
          	
          	\begin{figure}
          		\centering
          		\includegraphics{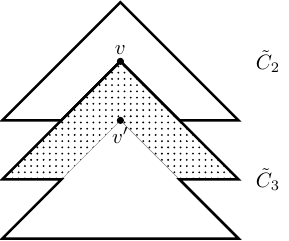}
          		\caption{
          			Construction of $\tilde{C}_1,\tilde{C}_2,\tilde{C}_3$
          			in the proof of Proposition~\ref{proposition:unifConstruct4}.
          			The dotted part (to be pumped as $\tilde{C}_1$ between $\tilde{C}_2$ and $\tilde{C}_3$)
          			is replaced with the clique-decomposition $\tau'_i$
                                corresponding to representative $\tau_i$ equivalent to $\mathcal{C}(v)$,
          			with $i\in\int{s'}$.
          		}
          		\label{figure:qunifPump}
          	\end{figure}
          	
          	Recall that $t(\tilde{\mathcal{C}_3}) = t(\mathcal{C}(v)) = t(\mathcal{C}(v'))$, 
          	and since  $t(\mathcal{C}(v)) = t(\tau_i)$ all %the graphs generated by
          	these four clique-decompositions have the same type $t$.
          	Moreover, let $u$ the node in $\tau_i$ such that $\tau_i = \tau'_i \oplus \tau_i(u)$
          	and $t(\tau_i) = t(\tau_i(u))$.
          	Hence, considering the map $\Lambda_{\tau'_i}$ from Lemma~\ref{lem:compositionality-oplus}, we have
          	$\Lambda_{\tau'_i}(t) = t$. 
          	Thus, gluing any of the four clique-decomposition having type $t$
                %that are generated by 
          	%$\tau(u), \tau_i, \mathcal{C}(v')$ and $\tilde{\mathcal{C}_3}$
                to $\tau'_i$
          	will give a graph of type $t$.
          	In particular, since $\tilde{\mathcal{C}_1} = \tau'_i$, the type of %the graph generated by 
          	$\tilde{\mathcal{C}_1} \oplus \dots \oplus \tilde{\mathcal{C}_1} \oplus \tilde{\mathcal{C}_3}$ is 
          	$t$, whatever the number of repetitions of $\tilde{\mathcal{C}_1}$. Finally, 
          	since $\mathcal{C}= \tilde{\mathcal{C}_2} \oplus \tilde{\mathcal{C}_3}$ has type $t_0$, 
          	we have $\Lambda_{\tilde{\mathcal{C}_2}}(t) = t_0$ and the type of 
          	$\tilde{\mathcal{C}_2} \oplus \tilde{\mathcal{C}_1} \oplus \dots \oplus \tilde{\mathcal{C}_1} \oplus \tilde{\mathcal{C}_3}$ 
          	is always $t_0$, which concludes the proof because $\phi\in t_0$.
        \end{proof}
        
        %%%%%%%%%%%%%%%%
        \subsection{Assembling $q$-uniform networks}
        
        We prove Theorem~\ref{theorem:unifMainMSO} for any $q\geq 2$,
        based on arithmetical considerations generalizing the case $q=2$
        presented in Subsection~\ref{ss:proof2}.
        The proof structure consists in applying Proposition~\ref{proposition:unifConstruct4}
        to obtain three clique-decompositions of width at most $k$, then incorporate the saturating graph $\Omega_m$,
        and finally reduce from \textbf{SAT} or \textbf{UNSAT} by constructing automata networks
        whose sizes are powers of $q$, thanks to the arithmetics from Theorem~\ref{theorem:interinfQ}.

        One may remark from the condition of Proposition~\ref{proposition:unifConstruct4}
        on the size of $\tilde{\mathcal{C}}_2\oplus\tilde{\mathcal{C}}_3$
        that the case $\ell=0$ will be important,
        and justifies its consideration in Section~\ref{section:lowerbounds}.
        It is required by our reasoning, in order to control the size of the initial solution
        to $\akbq{a}{b}{q}$.

        The other key ingredient consists in replacing the pumped part by its representative among
        $\{\tau'_i\}_{i\in\int{s'}}$, which allows to bound the value of $a$
        and apply Proposition~\ref{proposition:unifConstruct4} with a value of $x$
        suitable for the subsequent use of the equivalent condition given by
        Theorem~\ref{theorem:equivCond} for the application of Theorem~\ref{theorem:interinfQ}.

        \begin{proof}[Proof of Theorem~\ref{theorem:unifMainMSO}]
          The proof structure is analogous to Theorem~\ref{theorem:2NPhard}, %(for $q=2$),
          with additional considerations to apply the arithmetics from Section~\ref{section:lowerbounds}.
          The deterministic case is handled with the addition of $\chi$ as in Section~\ref{section:det},
          in order to have graphs of out-degree $1$ all the way.

          Let $q\geq 2$, and $\psi$ be a $q$-cw-non-trivial (or $q$-non-trivial)
          MSO formula of rank $m$.
          Without loss of generality, assume that the saturating graph
          $\Omega_m$ from Proposition~\ref{proposition:saturatingGraph} (or Corollary~\ref{corollary:saturatingGraph})
          is a counter-model of $\psi$,
          and let us build a reduction from \textbf{UNSAT}
          (otherwise consider $\neg\psi$ instead of $\psi$,
          and a reduction from \textbf{SAT} instead of \textbf{UNSAT}).

          Before calling Proposition~\ref{proposition:unifConstruct4},
          we setup an appropriate value of $x$ to ensure that the condition of Theorem~\ref{theorem:equivCond}
          will be verified.
          It is related to the value of $a$ which is the size of the glued part,
          \emph{i.e.}~$\tau'_i$ from the Proposition.
          %Recall that $\oplus$ merges the ports, of size $k$ in our context.
          Let $\eta$ be a value greater or equal to the maximum coprime power of $q$
          for $\{|\tau'_i| \mid i\in\int{s'}\}$,
          and also greater or equal to $\max\{\log_q(|\tau'_i|) \mid i\in\int{s'}\}$.

          Applying Proposition~\ref{proposition:unifConstruct4} for $x=\eta$,
          we obtain three $k$-clique-decompositions $\tilde{\Gamma}=\{\tilde{\mathcal{C}}_1,\tilde{\mathcal{C}_2},\tilde{\mathcal{C}}_3\}$,
          with $\tilde{\mathcal{C}}_1=\tau'_i$ for some $i\in\int{s'}$,
          such that $\Delta^{\tilde{\Gamma}}(2\cdot 1^\ell\cdot 3)\models\psi$ for all $\ell\in\N$.
          We deduce the existence of $(a,b)\in\N_+\times\N$ with:
          \begin{itemize}[nosep]
            \item $b=|\tilde{\mathcal{C}}_2\oplus\tilde{\mathcal{C}}_3|=q^N$ for some $N\geq\eta$,
            \item $a=|\tilde{\mathcal{C}}_1|=|\tau'_i|$.
          \end{itemize}
          We have $|\Delta^{\tilde{\Gamma}}(2\cdot 1^\ell\cdot 3)|=a\ell+b$,
          which is a power of $q$ for $\ell=0$,
          \emph{i.e.}~we have the initial solution $(0,N)$ to $\akbq{a}{b}{q}$.
          Since this solution verifies that the coprime power of $q$ for $a$
          is smaller than $N$, by Theorem~\ref{theorem:equivCond} there exists a couple
          $(\mu,\kappa)\in\N_+\times\N_+$ such that we can apply Theorem~\ref{theorem:interinfQ},
          and deduce that $\akbq{a}{b}{q}$ contains a geometric sequence of integers.
          More precisely, from the proof of Lemma~\ref{lemma:interinfQ},
          in $\Delta^{\tilde{\Gamma}}(2\cdot 1^\ell\cdot 3)$ we have graphs of size
          $q^{N+\ell'\mu}=bq^{\ell'\mu}$ for all $\ell'\in\N$,
          which correspond to $\ell=\frac{bq^{\ell'\mu}-b}{a}$ being an integer
          (the number of copies of $\tilde{\mathcal{C}}_1$).

          Before continuing, let us consider the fourth graph.
          Let $\alpha$ be the least power of $q$ such that
          $\mathcal{C}_1=\bigoplus_{i=1}^{\alpha}\tilde{\mathcal{C}}_1$ verifies $|\mathcal{C}_1|\geq|\Omega_m|$,
          and let $\mathcal{C}_0$ be the disjoint union of $\mathcal{C}_m$
          (a clique-decomposition of $\Omega_m$)
          and $|\mathcal{C}_1|-|\Omega_m|$ constant$^\circ$ nodes,
          so that $|\mathcal{C}_0|=|\mathcal{C}_1|$.
          Finally, let $\mathcal{C}_2=\tilde{\mathcal{C}}_2$ and $\mathcal{C}_3=\tilde{\mathcal{C}}_3$.
          The clique-decompositions %$\Gamma'=\{G_0,G_1,G_2,G_3\}$ admits clique-decompositions 
          $\Gamma=\{\mathcal{C}_0,\mathcal{C}_1,\mathcal{C}_2,\mathcal{C}_3\}$ 
          verify Conditions \emph{(i)--(iii)} of Proposition~\ref{pro:sat}.
          Furthermore, $\mathcal{C}_1$ is made of $\alpha$ copies of $\tilde{\mathcal{C}}_1$.

          Given a propositional formula $S$ on $s$ variables,
          we need at least $2^s$ copies of $\mathcal{C}_0$ or $\mathcal{C}_1$ in order to implement the reduction from \textbf{UNSAT},
          where for each valuation the circuit of the AN (resp.~NAN) builds
          a copy of the graph generated by $\mathcal{C}_0$ if $S$ is satisfied,
          and a copy of the graph generated by $\mathcal{C}_1$ otherwise.
          Recall that $\mathcal{C}_1$ is made of $\alpha$ copies of $\tilde{\mathcal{C}}_1$
          and $|\mathcal{C}_0|=|\mathcal{C}_1|$, that is, for $|w|=2^s$
          we have $|\Delta^\Gamma(2\cdot w\cdot 3)|=a\alpha 2^s+b$.
          The goal is now to find a suitable padding by $L(s)$ copies of $\tilde{\mathcal{C}}_1$,
          \emph{i.e.}~such that:
          \[
            |\Delta^{\Gamma'}(2\cdot w \cdot 4^{L(s)}\cdot 3)|=a(\alpha 2^s+L(s))+b
          \]
          is a power of $q$, with $|w|=2^s$, $\mathcal{C}_4=\tilde{\mathcal{C}}_1$
          and $\Gamma'=\{\mathcal{C}_0,\mathcal{C}_1,\mathcal{C}_2,\mathcal{C}_3,\mathcal{C}_4\}$.
          Then the reduction is as usual: the graph generated by $\Delta^{\Gamma'}(2\cdot w \cdot 4^{L(s)}\cdot 3)$
          is a model of $\psi$ if and only if $w$ does not contain letter $0$ 
          (in this case it has only copies of $\tilde{\mathcal{C}}_1$).
          Remember that the only non-constant part is $S$ (and $s$).

          We want to find $\ell'\in\N$ such that:
          \[
            \alpha 2^s+L(s) = \frac{bq^{\ell'\mu}-b}{a}
            \iff
            L(s) = \frac{b}{a}(q^{\ell'\mu}-1)-\alpha 2^s.
          \]

          A solution with $L(s)\geq 0$ is $\ell'=s+\log_q(\alpha)+1$.
          Indeed, we have $\frac{b}{a}\geq 1$ by the hypothesis on $\eta$ (upper bounding $a$) and $b=q^\eta$.
          Then $q^{\ell'\mu} \geq \alpha 2^s+1$ is verified by $\ell'$ because $\mu\geq 1$ and
          $q\geq 2$\footnote{$\ell'=\left\lceil\frac{\log_q(\frac{a}{b}\alpha 2^s+1)}{\mu}\right\rceil$
          would be a smaller valid solution, and
          $\alpha$ does not need to be a power of $q$.}.

          Therefore, we can build the circuit of a $q$-uniform NAN (or AN) with
          \begin{align*}
            n=\log_q(b)+\mu(s+1)+\mu\log_q(\alpha) &\quad\text{automata, and}\\
            \alpha 2^s+L(s)=\frac{bq^{\mu(s+1)}\alpha^\mu-b}{a} &\quad\text{copies of $\tilde{\mathcal{C}}_1$}
          \end{align*}
          (some copies of $\tilde{\mathcal{C}}_1$ form copies of $\mathcal{C}_1$, or are replaced by a copy of
          $\mathcal{C}_0$ when $S$ is satisfied).
          Recall that $a$, $b$, $\mu$ and $\alpha$ are constants depending only on $\psi$.
        \end{proof}

        %%%%%%%%%%%%%%%%%%%%%%%%%%%%%%%%
        \section{Conclusion and perspectives}
        \label{section:perspectives}

        We have proven Theorem~\ref{theorem:hard}, that any non-trivial (in the deterministic setting)
        or cw-non-trivial (in the non-deterministic setting) MSO question on the dynamics of a given
        automata network (AN) or non-deterministic automata network (NAN)
        is $\NP$-hard or $\coNP$-hard to answer, when given the network's description
        in the form of a circuit, even in the case of bounded alphabets.
        Part~c of Theorem~\ref{theorem:hard} was proven in~\cite{gglgopt25},
        on which our developments are grounded.
        In Section~\ref{section:det}, we have included the deterministic constraint to the
        abstract pumping technique in order to obtain Theorem~\ref{theorem:detMainMSO}.
        In Section~\ref{section:lowerbounds} we have treated the arithmetics of the intersection
        between the sizes of $q$-uniform automata networks, and the sizes of models obtained by pumping.
        This led to the proof of Theorem~\ref{theorem:unifMainMSO} in Section~\ref{section:qunif}.
        In particular, the Rice-like complexity lower bounds also hold for
        Boolean (deterministic and non-deterministic) automata networks, a model widely used
        in applications to system biology, for example in the study of gene regulatory networks.
        Our results intuitively state that \emph{any} ``interesting'' computational investigation
        of the dynamics in a model of interacting entities, is limited by
        a ``high'' time-complexity cost in the worst case.
        Hence there cannot exist any general efficient algorithm,
        unless some unsuspected complexity collapse occurs (such as $\Poly=\NP$).
        %recall the results, and that they indeed complete the proof of Theorem~\ref{theorem:hard}
        %recall the importance of the results for the concrete algorithmic study (by the mean of computers,
        %a widespread approach nowadays) of natural models of computation.
	
        The deterministic case is now closed for MSO (remark that deterministic implies cliquewidth at most $6$), 
        because the trivial versus non-trivial dichotomy is sharp and deep:
        all trivial formulas are answers with $\O(1)$ algorithms,
        whereas all non-trivial formulas are hard for the first level of the polynomial hierarchy
        (either $\NP$ or $\coNP$).
        However, in the non-deterministic case there are computationally hard
        cw-trivial formulas (\emph{e.g.}~being a grid), hence the dichotomy does not
        reach the sharpness of an ``a la Rice'' statement.
        To go further, it would be meaningful to extend the proof technique to other structural graph parameters,
        such as rankwidth and twinwidth.
        In~\cite{gglgopt25}, it is formally argued that graph parameterization is necessary
        to derive general complexity lower bounds for non-deterministic automata networks,
        because a proof of the basic trivial versus non-trivial dichotomy in this setting
        implies unexpected complexity collapses (regardless of our considerations on alphabet sizes,
        and even at first-order).

        The definition of non-determinism we have employed is \emph{global},
        and in the uniform case it does not accommodate the Cartesian product of local possibilities
        (in the case of unrestricted alphabets it does, simply by considering
        automata networks of size $n=1$, with potentially huge alphabets).
        Could the abstract pumping technique be applied to such a notion of local non-determinism?

        A result of~\cite{ggpt21} is that there are first order questions complete at all levels of
        the polynomial hierarchy (even in the deterministic case),
        therefore obtaining general lower bounds at higher levels
        seems another pertinent track to follow,
        where a good notion of ``highly non-trivial'' formula needs to be established.

        Finally, the signature of our logics contains (apart from the ``$\in$'' of MSO)
        only the two binary relations $\{=,\to\}$ on pairs of configurations (vertices),
        consequently all expressible questions are valid up to graph isomorphism.
        Distinguishing configurations may be meaningful for applications,
        therefore the addition of finer relations is another interesting research track
        to expand the results.

        %%%%%%%%%%%%%%%%%%%%%%%%%%%%%%%%
        \section*{Acknowledgments}

        The authors are thankful to Colin Geniet, Guilhem Gamard and Guillaume Theyssier
        for helpful and stimulating discussions on technical and non-technical aspects of this work.
        They also thank the support of projects
        ANR-24-CE48-7504 ALARICE,
        HORIZON-MSCA-2022-SE-01 101131549 ACANCOS and
        STIC AmSud CAMA 22-STIC-02.

\bibliographystyle{abbrv} 
\bibliography{biblio}

\end{document}